\documentclass[12pt,preprint]{aastex}
\usepackage{amssymb}

\usepackage{amsmath}

\bibpunct{\hspace{-4pt}}{}{}{a}{}{}
\makeatletter

\begin{document}

\title{The Stellar Cusp Around the Supermassive Black Hole in the Galactic
Center }

\author{R.Genzel\altaffilmark{1}, R. Sch\"{o}del, T.Ott, F.Eisenhauer, R.
Hofmann, M. Lehnert}
\affil{Max-Planck Institut f\"{u}r extraterrestrische Physik, Garching,
Germany}

\and

\author{A. Eckart}
\affil{1.Physikalische Institut, Universit\"{a}t K\"{o}ln, K\"{o}ln, Germany}

\and

\author{T. Alexander}
\affil{Faculty of Physics, Weizmann Institute of Science, Rehovot, Israel}

\and

\author{A. Sternberg}
\affil{School of Physics and Astronomy, Tel Aviv University, Ramat Aviv, Israel}

\and

\author{R.Lenzen}
\affil{Max-Planck Institut f\"{u}r Astronomie, Heidelberg, Germany}

\and

\author{Y.Cl\'enet, F. Lacombe, D. Rouan}
\affil{Observatoire de Paris-Meudon, Meudon, France}

\and

\author{A.Renzini, L.E. Tacconi-Garman}
\affil{European Southern Observatory, Garching, Germany}

\altaffiltext{1}{also Department of Physics, University of California, Berkeley, USA}

\begin{abstract}
We analyze deep near-IR adaptive optics imaging (taken with NAOS/CONICA on
the VLT) \footnote{
Based on observations obtained at the European Southern Observatory, Chile}
as well as new proper motion data of the nuclear star cluster of the Milky
Way. The surface density distribution of faint (H$\leq20$, K$_{s}\leq19$)
stars peaks within 0.2$^{\prime\prime}$ of the black hole candidate SgrA$%
^\star $. The radial density distribution of this stellar 'cusp' follows a
power law of exponent $\alpha\sim 1.3-1.4$. The K-band luminosity function
of the overall nuclear stellar cluster (within $9^{\prime\prime}$ of SgrA$%
^\star $) resembles that of the large scale, Galactic bulge, but shows an
excess of stars at K$_{s}\leq14$. It fits population synthesis models of an
old, metal rich stellar population with a contribution from young, early and
late-type stars at the bright end. In contrast, the cusp within $\leq
1.5^{\prime\prime}$ of SgrA$^\star$ appears to have a featureless luminosity
function, suggesting that old, low mass horizontal branch/red clump stars
are lacking. Likewise there appear to be fewer late type giants. The
innermost cusp also contains a group of moderately bright, early type stars
that are tightly bound to the black hole. We interpret these results as
evidence that the stellar properties change significantly from the outer
cluster ($\geq$ a few arcsecs) to the dense innermost region around the
black hole.

We find that most of the massive early type stars at distances 1-10" from
SgrA$^\star$ are located in two rotating and geometrically thin disks. These
disks are inclined at large angles and counter-rotate with respect to each
other. Their stellar content is essentially the same, indicating that they
formed at the same time. We conclude that of the possible formation
scenarios for these massive stars the most probable one is that 5-8 million
years ago two clouds fell into the center, collided, were shock compressed
and then formed two rotating (accretion) disks orbiting the central black
hole. For the OB-stars in the central arcsecond, on the other hand, a
stellar merger model is the most appealing explanation. These stars may thus
be 'super-blue-stragglers', formed and `rejuvenated' through mergers of
lower mass stars in the very dense ($\geq 10^{8}\,M_{\odot }$pc$^{-3})$
environment of the cusp. The `collider model' also accounts for the lack of
giants within the central few arcseconds.

The star closest to SgrA$^\star$ in 2002, S2, exhibits a 3.8 $\micron$
excess. We propose that the mid-IR emission either comes from the accretion
flow around the black hole itself, or from dust in the accretion flow that
is heated by the ultra-violet emission of S2.
\end{abstract}


\section{Introduction}

Because of its proximity (distance 8 kpc, Reid 1993), the center of the
Milky Way is a unique laboratory for studying the physical processes in
galactic nuclei (e.g. Genzel, Hollenbach \& Townes 1994, Morris \& Serabyn
1996, Mezger, Duschl \& Zylka 1996, Melia \& Falcke 2001). In particular,
the Galactic Center offers the unique opportunity for investigating stars
and gas in the immediate vicinity of a supermassive black hole, at a level
of detail that will not be accessible in any other galactic nucleus in the
foreseeable future. In the present paper we present and analyze new high
resolution, adaptive optics imaging and observations of stellar velocities
in the nuclear cluster, relating to its spatial distribution, evolution and
dynamics. Let us summarize what is known from observations during the past
decade.

First, the overall surface density and surface brightness of stars increases
with decreasing separation from the center near the compact radio source SgrA%
$^{\star }$ (diameter $\sim $10 light minutes, Doeleman et al. 2001), with a
near-isothermal power law (stellar volume density $\varpropto $ $R^{-2}$)
from projected radii $p\leq 100^{\prime \prime }$ to about $10^{\prime
\prime }$ (Catchpole, Whitelock \& Glass 1990). Within that radius the
surface brightness continues to increase inward to \symbol{126}1'' but the
stellar surface number density at moderately bright magnitudes (K$\leq 15$ 
\footnote{
characteristic of main sequence stars earlier than B0, and giants later than
K4}) flattens, consistent with a half-peak surface brightness/density (core)
radius of $\sim $0.34 $\pm $ 0.2 pc (Genzel et al. 1996). The value of the
core radius is uncertain. The lower limit comes from surface density counts
of the brightest few hundred stars (Eckart et al. 1993). The upper limit
comes from the surface brightness distribution of late type stars (Allen
1994), and measurements of the integrated light of faint stars in between
the brightest stars (Rieke \& Rieke 1994). The largest concentration of K$%
\leq 13$ stellar light and surface density is not centered on SgrA$^{\star }$%
, but on the IRS16 complex about 2$^{\prime \prime }$ east of it. This
offset has been interpreted as evidence that SgrA$^{\star }$ cannot be a
massive black hole (e.g. Allen \& Sanders 1986).

Second, there are several different stellar populations/components in the
central parsec (for a review see Genzel 2001). The stellar mass and the
near-IR light at K$\geq 13$ is dominated by red giants in the old (1-10 Gyr)
component of the nuclear star cluster. A group of about a dozen luminous,
blue supergiants (`HeI emission line stars') strongly affects the near-IR
maps at the bright end (K$\sim 9-12$), and probably indicates recent
formation of massive stars within the last 2-7 Myrs (Forrest et al. 1987,
Allen, Hyland \& Hillier 1990, Krabbe et al. 1991, 1995, Tamblyn et al.
1996, Blum et al. 1996b, Paumard et al. 2001). A number of bright (K$\sim
10-12$) asymptotic giant branch (AGB) stars sample an intermediate mass,
intermediate age component ($\geq $100 Myr, Lebofsky \& Rieke 1987, Krabbe
et al. 1995, Blum et al. 1996b). Finally there is a group of dust embedded
stars with near-featureless near-IR spectra (Becklin et al. 1978, Krabbe et
al. 1995, Genzel et al. 1996), many of which are associated with the gaseous
mini-spiral (Lo \& Claussen 1983). Their nature is uncertain. The near-IR
luminosity function to K$\leq $16 resembles that of the Galactic Bulge in
Baade's Window (Blum et al. 1996a, Davidge et al. 1997, Alexander \&
Sternberg 1999).

Third, the mean stellar velocities (or velocity dispersions) follow a Kepler
law ($\langle v^{{\ 2}}\rangle \propto R^{{\ -1}}$) from $\sim 0.1^{\prime
\prime }$ to $\geq 20^{\prime \prime }$, and provide compelling evidence for
the presence of a central compact mass (Genzel et al. 1996, 97, 2000; Eckart
and Genzel 1996, 97; Ghez et al. 1998). Overall the stellar velocities are
consistent with an isotropic velocity field but the HeI emission line stars
appear to be preferentially on tangential orbits (Genzel et al. 2000).
Statistical modeling of the stellar velocity data (projected mass
estimators, Jeans equation modeling, non-parametric modeling) imply that the
2 to 3.3$\times 10^{6}\,M_{\odot }$ central dark mass is concentrated on
scales less than ten light days (Eckart \& Genzel 1996, 1997, Ghez et al.
1998, 2000, Genzel et al. 2000, Chakrabarti \& Saha 2001, Eckart et al.
2002). Most recently, Sch\"{o}del et al. (2002, 2003) and Ghez et al. (2003)
found that the star closest to SgrA$^{\star }$ in 2002, S2, is on a highly
elliptical Keplerian orbit around the radio source. The enclosed mass within
the orbit's peri-center approach of 17 light hours (in April 2002: at which
point the star had a sky projected velocity of $\geq $5000 km/s) is $3.5\pm
0.5\times 10^{6}\,M_{\odot }$ (Sch\"{o}del et al. 2003), which is within the
uncertainties the same as the mass at distances of ten to several hundred
light days. As a result the density of a hypothetical, non-black hole
configuration has to have a central density of 2$\times 10^{17}\,M_{\odot }$
pc$^{-3}$ or greater. When combined with the 8 to 20 km/s upper limit to the
proper motion of SgrA$^{\ast }$ (Backer \& Sramek 1999, Reid et al. 1999,
Reid et al. 2003b) and a theoretical analysis of the motion of a black hole
in the central star cluster, the density limit increases to $\geq
10^{19.5}\,M_{\odot }$ pc$^{-3}$ (Sch\"{o}del et al. 2003). The new orbital
data now definitely exclude a dark cluster of astrophysical objects
(e.g.neutron stars), or a ball of 10-60 keV fermions as possible
configurations of the central mass concentration. The only non-black hole
configuration is a ball of hypothetical, heavy bosons, which would not be
stable, however. The gravitational potential in the central light year of
the Galactic Center thus is almost certainly dominated by a massive black
hole associated with SgrA$^{\star }$.

We would like to better determine the distribution, properties and dynamics
of the different stellar components and understand their formation and
evolution. We need to sample the numerous lower mass stars in the nuclear
star cluster to determine the overall stellar density distribution and
centroid. Theory predicts a number of effects that are unique to the
environment of a black hole (for a review see Alexander 2002). There should
be a density `singularity' of the star cluster (a `stellar cusp') centered
on the hole (Bahcall \& Wolf 1976, 1977, Young 1980). Because of the very
dense environment close encounters and direct stellar collisions can occur
and alter the stellar distribution, dynamics and population near the hole.
Giant stars can be destroyed (Davies \& Benz 1991, Rasio \& Shapiro 1990,
Davies et al. 1998, Alexander 1999) and exotic stars created (e.g. Thorne
and Zytkow 1975). Given the extensive evidence for hot, massive stars in the
central parsec we need to answer how they came into being in the very dense
central environment. We also would like to detect SgrA$^{\star }$ itself in
the near-/mid-IR, and place tighter constraints on the accretion flow
properties (Melia \& Falcke 2001). We will address all these issues in the
present paper.

Previous data sets from the 3.5m ESO New Technology Telescope (NTT: Eckart
\& Genzel 1996, 1997, Genzel et al. 1997, 2000), the 3.6m Canada France
Hawaii telescope (CFHT: Davidge et al. 1997), and from the 10m Keck
telescope (Ghez et al. 1998, 2000) have FWHM resolutions of 0.15$^{\prime
\prime }$, 0.16$^{\prime \prime }$ and 0.05$^{\prime \prime }$ respectively,
and reach to K$\leq $16-17. Deeper imaging at the diffraction limited
resolution of 8-10m class telescopes is critical for addressing the problems
outlined above. For this purpose we have carried out deep near-infrared
adaptive optics (AO) observations with the new NAOS/CONICA instrument on the
ESO VLT (Lenzen et al. 1998, Rousset et al. 1998, Brandner et al. 2002).
These observations provide diffraction limited images at H (1.65$\mu m)$ and
K$_{s}$ (2.16$\mu m)$ to $\sim 20$ and at L$^{\prime }$(3.76$\mu m)$ to $%
\sim 14$. These improvements are largely due to the near-infrared wavefront
sensor of NAOS (which can lock on the bright star IRS7 within a few arcsec
of SgrA*) and due to the fact that the Galactic center passes through the
zenith at Paranal. As a result NAOS/CONICA achieves diffraction limited
imaging with stable and high (40-50\%), K-band Strehl ratios. Proper motions
and radial velocities for about 1000 stars are now available from a new
analysis of the ESO-NTT proper motions and integral field spectroscopy
between 1992 and 2003 (Ott et al. 2003). We combine the AO observations with
the new data base of stellar proper motions and radial velocities, for a new
analysis of the properties of the nuclear star cluster.

\section{Observations and Data Reduction}

The observations presented in this paper were carried out during science
verification with the NAOS/CONICA adaptive optics system/near-infrared
camera at the unit telescope 4 (Yepun) of the ESO VLT in Paranal, Chile.
H-band (1.65$\mu m)$, K$_{s}$-band (2.16$\mu $m) and $L^{\prime }$-band (3.76%
$\mu $m) imaging data were taken on August 29, 2002. Seeing was between 0.5
and 0.6$^{\prime \prime }$. For H\ and K$_{s}$ the infrared wavefront sensor
of NAOS was used to close the feedback loop on the bright supergiant IRS 7 $%
\sim 5.5^{\prime \prime }$ north of SgrA$^{\star }$. The detector pixel
scale was 0.0132$^{\prime \prime }$/pixel for H and K$_{s}$ and 0.027$"$%
/pixel for L$^{\prime }$. The diffraction limited resolution was 40, 55 and
95 milliarcseconds (mas) in the H, K$_{s}$ and L$^{\prime }$ bands (FWHM).
The oversampling helped to reduce the saturation of the numerous bright
sources in the nuclear cluster. The detector integration time was 15 s. Four
individual exposures were combined by a pipeline into one image of 60 s
integration time. 25 such images were taken in the H band and 20 in the K$%
_{s}$-band. The individual images were flat-fielded, sky-subtracted and
corrected for dead/bad pixels. The final frames were co-added with a simple
shift-and-add (SSA) algorithm to final images of 1200 s (K$_{s}$ band,
Figure 1) and 1500 s (H band, Figure~2) total integration time. The Strehl
ratio measured on individual sources near the guiding star is $\sim 50\%$ in
the K$_{s}$-band and $\sim 33\%$ in the H band. Dithering between the
exposures resulted in a field-of-view (FOV) larger than the $14\times
14^{\prime \prime }$ provided by CONICA with the finest pixel scale. From
the final mosaic we selected a $\sim 9^{\prime \prime }\times 9^{\prime
\prime }$ region with the highest S/N-ratio, approximately centered on SgrA$%
^{\star }$. There are numerous ($\sim 40-50$) bright (K$_{s}<12$) stars in
the FOV, which are strongly saturated in the K$_{s}$-band image. Saturation
is less severe in the H-band image. For L' the AO loop was closed with the
visible light wavefront sensor on a $V\sim 14$ star $\sim 20^{\prime \prime
} $ to the north-east of SgrA$^{\star }$ (at the time of the observations
the dichroic allowing the simultaneous use of the infrared wavefront sensor
and the L/M-bands was not available). A single imaging frame consists of 150
exposures of 0.2 s, coadded by a pipeline into a single frame with 30 s
integration time. 76 such images were taken. After flat fielding,
sky-subtraction and dead/bad pixel correction, the individual frames were
combined into a final SSA image of 2280 s integration time, with a Strehl
ratio of $\sim 50\%$ (see also Cl\'{e}net et al. 2003). The right panel of
Figure 1 shows a color composite of all three images.

\bigskip

In the following subsections we discuss in detail the analysis and methods
we have developed and applied to carry out source counts, photometry and
astrometry in our crowded, high dynamic range, adaptive optics images.
Readers who wish to skip these technical details should continue with
section 3.

\bigskip

\subsection{Source identification and photometry}

In order to facilitate the identification of sources in the extremely dense
nuclear stellar cluster and to reduce the influence of the seeing halos of
the numerous bright stars, we deconvolved the images prior to number counts
and photometric analysis. A point-spread function (PSF) was extracted from
the images by taking the median of several (more than 10 in the case of the K%
$_{s}$ and H-band images) fairly isolated, bright, but not saturated stars.
We deconvolved the images using two methods, a linear Wiener-filter
technique (Ott, Eckart \& Genzel 1999), and the Lucy-Richardson (LR: Lucy
1974) algorithm. The delta maps resulting from the LR deconvolution were
re-convolved with a Gaussian PSF of the appropriate FWHM for the respective
wavelengths. For comparison of the results of the different image processing
techniques, Figure 2 shows the direct (SSA) image, and the
Wiener-filtered/linearly deconvolved and the Lucy-deconvolved H-band images
of the central $\sim 2^{\prime \prime }$ around the compact radio source SgrA%
$^{\star }$. The agreement of sources identified with the different image
analysis techniques is generally very good. Only within a few tenths of an
arcsec of bright (mostly saturated) stars some deterioration is caused by
artefacts such as ringing. In these regions, graininess of the seeing halo,
ringing and streaks make source identification of stars 4 or more magnitudes
fainter than the bright star unreliable. Figure 3 shows H-, K$_{s}$- \ and L$%
^{\prime }$-band, Lucy-deconvolved images of the central region around SgrA$%
^{\star }$.

We identified point sources and carried out photometry with the FIND
procedure from the IDL Astrolib library. The IDL FIND procedure convolves
the image with a gaussian beam prior to searching for local maxima. We found
that there is no ideal choice for this parameter. Depending on the exact
value of the FWHM parameter, sources may be identified or missed in densely
packed regions, while various spurious sources may be detected. We therefore
decided to repeat the source detection (and completeness correction, see
below) procedure three times, with different values of the FWHM parameter.
For construction of the final source lists we compared the lists of sources
from the LR and Wiener deconvolved images, eliminating sources that only
showed up in one of the lists. For the remaining stars, their photometric
values were averaged, with the error taken as the deviation of the two
measurements from the average. The photometric errors for the different
bands are shown in Figure 4. In regions not too close to bright stars,
sources of 19th magnitude are $\sim 5\sigma $ above the background in the
deconvolved K$_{s}$-band image, while the H -band image is about 1.5
magnitudes deeper. Spurious sources are detected in the seeing halos of
bright stars. This problem was particularly important for the K$_{s}$-band,
where numerous bright stars were strongly saturated. Deconvolution
techniques also tend to create faint, spurious point sources. We therefore
adopted a conservative approach by applying the constraint that genuine
stellar sources must be present in both the H- and K$_{s}$-band image. We
verified that this procedure effectively excluded spurious detections by
comparing the maps of the identified stars with the actual images.

The final lists of sources comprise between 3200 and 4000 stars, depending
on the choice of the FWHM parameter, with significantly different source
counts only for sources with K$_{s}\geq $18. After correction for
incompleteness (see below), however, the number counts at all magnitudes
agree very well. For the number counts of the luminosity function and of the
surface number density presented in this paper, we chose the average of the
counts resulting from the different choices of the FWHM parameter. For the
errors, we combined quadratically the statistical counting (Poisson) error
and the maximum deviation of the counts from their average, which
corresponds to a $\sim $10\% uncertainty at K$_{s}\leq 19$. We calibrated
the photometry in the H, K$_{s}$ and L$^{\prime }$ bands relative to sources
in the cluster. The main limitation in identifying appropriate calibration
stars resulted from finding stars that were not saturated, but bright enough
to be included in the Blum et al. (1996a) list. Additionally, these stars
should be fairly isolated because the Blum et al. (1996a) data were seeing
limited. By considering these points we tried to minimize possible
systematic errors in flux calibration. For the K$_{s}$-band image, we chose
four isolated stars with magnitudes between 13 and 14 from the Ott et al.
(2003) list ( IDs 412, 284, 265, and 239). For the H-band photometry we used
the Blum et al. (1996) photometric value for IRS 33N (named IRS A11 in their
list). We took IRS 33N and IRS 16CC (Blum et al. 1996a) to calibrate the L$%
^{\prime }$-band data. Another point to keep in mind is that the Blum et al.
(1996a) data are made for the K and L-bands, while the observations
presented in this paper were made with L$^{\prime }$ and K$_{s}$ band
filters. We estimate the resulting relative uncertainties of our photometry
to be less than 0.1 mag at K$_{s}\leq $18 (H$\leq $19), while the absolute
photometry is uncertain by 0.15 mag in H and K$_{s}$, and 0.3 mag at L$%
^{\prime }$.

\subsection{Incompleteness correction}

We determined the incompleteness correction for the K$_{s}$-band images with
the well known technique of first adding and then again recovering
artificial stars. Taking the same PSF as used for deconvolution, we inserted
artificial stars randomly into the original stellar field. The image
containing the artificial stars was Wiener deconvolved, followed by source
identification with the FIND procedure. We did not repeat this procedure
with a LR deconvolution because of the enormous amount of computational time
needed for this method. The artificial stars were spaced at intervals of $%
\sim 0.5^{\prime \prime }$, such that their individual PSFs did not
interfere with each other. By repeating the same procedure many times with
different positions for the artificial stars, we probed the image with
artificial stars in a dense $\sim 0.13^{\prime \prime }\times 0.13^{\prime
\prime }$ grid. We recorded in 'completeness maps' the probability of
recovering a source with a given magnitude at a given position. The
completeness map for stars of 18th magnitude in the K$_{s}$-band image is
shown in Fig.5. Its spatial structure of course reflects the distribution of
bright stars in Fig.1. Table 1 gives the overall completeness as well as the
average magnitude and standard deviation of the magnitude of the recovered
artificial stars for each magnitude interval. We determined the completeness
corrections for all three choices of the FIND FWHM parameter that were used
to create the source lists.

\subsection{Absolute astrometry}

In order to obtain astrometric positions for the stars relative to Sgr A*,
we aligned the infrared images, in which the stars are observed, with the
astrometrically accurate radio images, in which SgrA$^{\star }$ is observed
(Sch\"{o}del et al. 2002, Reid et al. 2003a, Ott et al.2003). For this
purpose we aligned our NAOS/CONICA images with an astrometric grid using all
5 to 7 SiO maser sources in the field of view whose positions are known
through measurements with the VLA and the VLBA with accuracies of a few mas
(Reid et al. 2003a, circled in Fig. 1). The SiO masers originate in the
central 10 AU ($\sim $1 mas) of the circumstellar envelopes of bright red
giants and supergiants, which are also present on the infrared images. The
position of the radio source SgrA$^{\star }$ on the infrared image, as
determined by a transformation that also takes account of up to second order
terms (Reid et al. (2003a)), has a 3$\sigma $ uncertainty of $\pm $ 30mas
and is denoted by arrows in the images in Figures 1 and 2 and as small
crosses in Figure 3. The astrometric position of SgrA$^{\star }$ is
coincident within a few milliarcsec with the `gravitational force center',
as determined from the orbit of S2 (Sch\"{o}del et al. 2002, 2003, Ghez et
al. 2003).

\subsection{Number counts and K- luminosity function}

We computed surface number densities by counting the stars in annuli with
increasing radius around Sgr A*. We then corrected these number counts by
dividing by the appropriate incompleteness at that radius and magnitude. In
order to avoid completeness corrections significantly larger than a factor
of 2 in the innermost annuli, where completeness and number counts are low
for faint stars, we only used stars with K-magnitudes brighter than 17 for
this analysis. In a second analysis we counted stars within $\sim 1"$ of Sgr
A* by eye in the Lucy-Richardson deconvolved H-Band image. This region is
devoid of bright stars and the H-band image has the advantage of lower
confusion. From the source counts, we also constructed completeness
corrected overall K-band luminosity functions (KLFs) for the circular
regions within $1.5"$ and within $9"$ of Sgr A*.

\subsection{Proper motion and narrow band data sets}

To study in more detail the dynamical, photometric and spectral properties
of the nuclear star cluster we combined our NAOS/CONICA imaging and
photometry with the new proper motion and radial velocity data base of Ott
et al. (2003). Ott et al. (2003) constructed their sample from 10 years
(1992-2002) of K-band speckle imaging data at the 3.5m ESO NTT, as well as
several H/K integral field spectroscopy data sets. It contains 881 stars
with proper motion errors $<$100 km/s, K$\leq $16 and projected offsets from
SgrA$^{\star }$ \ up to 12$".$ To separate between early and late type stars
in the broad-band maps at magnitudes fainter than the current K$\sim $13
spectroscopic limit, Ott et al. used publicly available narrow-band maps
taken as part of the Gemini North Galactic Center Demonstration Science Data
Set. From these maps Ott et al. (2003) constructed a photometric, overtone
CO-band index $m(\mathrm{CO)}$ for each star in the GEMINI data set, and
defined as 
\begin{equation}
m(\mathrm{CO})=m(2.29)-m(2.26)\,,
\end{equation}%
where $m(2.29)$ and $m(2.26)$ are the magnitudes in the narrow band filters
at 2.29 $\mu $m (CO v=0-2) and 2.26 $\mu $m (continuum), both calibrated on
the K-band magnitude scale. Figure 6 is a plot of $m(\mathrm{CO})$ as a
function of K for those 706 stars of the Ott et al. sample with $<$100 km/s
proper motion errors that also have GEMINI CO-indices. 86 of these stars
also have near-IR spectroscopic identifications and radial velocities
(Genzel et al. 1997, 2000, Paumard et al. 2001, Gezari et al. 2002, Ott et
al. 2003). The spectroscopic early type stars (HI and HeI emission lines and
lack of CO absorption bands) and late type stars (CO overtone absorption
bands) are marked as filled triangles and filled circles, respectively. The $%
m(\mathrm{CO})$-K plot shows that K$\leq 15$ early and late type stars can
be separated with good confidence at $m(\mathrm{CO})\sim 0.04$. \ Fainter
late type stars have weaker CO-indices, probably because they are early
K-giants. For these fainter stars it is not possible to distinguish between
early and late type stars on the basis of the GEMINI photometric index.

\section{Results}

\subsection{Multi-band imaging of the stellar cluster}

The NAOS/CONICA science verification H and K$_{s}$- images are $\sim 3$ mag
deeper than the earlier NTT and Keck images, the H-image is the highest
resolution image of the Galactic center region so far, and the L$^{\prime }$
image is the first one of the Galactic center at $\sim 0.1^{\prime \prime }$
resolution (for more details see Cl\'{e}net et al. 2003). Figures 1, 2 and 3
show our basic H, K$_{s}$ and L$^{\prime }$ data sets. We have chosen a
logarithmic color scale in all three figures to emphasize the large ( $\geq $%
13 mag) dynamic range and depth (H/K$_{s}\leq $20) of our H and K$_{s}$-
images. The faintest sources recognizable on the images are equivalent to $%
\sim $2 M$_{\odot }$ A5/F0 main sequence stars. The images in Figure 1
demonstrate the complexity of the dense stellar environment in the central
parsec. Bright blue supergiants (in the IRS16 and IRS13 complexes), as well
as red supergiants (IRS 7) and asymptotic giant branch stars (IRS12N, 10EE
and 15NE) dominate the H- and K$_{s}$-images. At L' there is an additional
group of dusty sources (IRS1W, 3, 21). Extended L' emission comes from hot
dust in the gaseous `mini-spiral' streamers comprising the most prominent
features of the SgrA West HII region.

The immediate vicinity of SgrA* lacks bright stars and dust. There is a
concentration of K$_{s}\geq $14 blue stars centered on the radio source (the
`SgrA* cluster'). A number of these blue 'S'-stars (Fig.3) are now known
from the proper motion studies to in fact reside in the central 25 light
days. The orbital parameters for 6 of these stars indicate that their are
bound to the central object with periods between 15 and a few hundred years
(Sch\"{o}del et al. 2002, 2003, Ghez et al. 2003).

\subsection{Spatial distribution: stellar cusp}

Figure 7 is a plot of the binned, stellar surface density distribution for
stars as a function of projected separation from SgrA$^{\star }$. As
described in section 2.2, we have corrected the observed K$_{s}\leq $17\
counts from Fig.1 for incompleteness. For the higher resolution H-band data
we plot the observed H$\leq $19.8 counts since the incompleteness correction
in the central 1.5'' is small and does not significantly vary with position.
We also have determined distributions to fainter levels (K$_{s}\leq $18),
with very similar results. However, the incompleteness corrections at those
fainter levels become large near the center and dominate the error budget.
The most reliable results are obtained for the magnitude limits taken in
Fig.7. For extrapolation to larger radii we combine the NAOS/CONICA data
with shallower (K$\leq 15$), but wider field NTT/SHARP number counts (Genzel
et al. 2000), appropriately scaled for the best match with the deeper
NAOS/CONICA data in the overlap region. As discussed in the Introduction,
the stellar counts for projected distances $p(\mathrm{SgrA}^{\star })\geq $
5-10'' can be reasonably well fit by a flattened isothermal sphere of core
radius $\sim $0.3 pc (Genzel et al. 1996 and references therein). Within a
few arcseconds the new data clearly indicate an excess of faint stars above
that of a flat core, already suggested by the earlier SHARP/NTT and Keck
data (Eckart et al. 1995, Alexander 1999). The surface density of faint
stars increases with decreasing separation from the radio source. The
smoothed, two dimensional distribution of faint stars in the H and K$_{s}$%
-images (Fig. 8) visually confirms the existence of this `cusp' and shows
that it is centered on SgrA$^{\star }$ [($\Delta \mathrm{RA},\Delta \mathrm{%
Dec}$)$=$($+0.09",-0.15"$)], within an uncertainty of $\pm $0.2$".$ This is
in contrast to the near-IR light distribution (Fig.1), which is centered on
the bright stars in the IRS16 complex. Our data thus resolve the 17 year old
puzzle of why SgrA$^{\ast }$ is offset from the 2$\mu m$ emission peak
(Allen \& Sanders 1986). The offset is caused by the bright stars in the
IRS16 complex, and is not a property of the majority of the faint stars in
the overall nuclear cluster.

Following Alexander (1999) we have analyzed the surface number density
distribution in Fig.7 with a broken power law, stellar density distribution,
with the simultaneous constraint that the (stellar) dynamical mass is 3.2,
8.4 and $27.3\times 10^{6}\,M_{\odot }$ at $R=$1.9, 3.8 and 11 pc
(subtracting from the mass distribution of Genzel et al. 1996 a $2.8\times
10^{6}\,M_{\odot }$ central point mass). The resulting fit shown in Fig. 7
has the following parameters, 
\begin{equation}
\rho _{\star }(R)=1.2\times 10^{6}\left( \frac{R}{10^{\prime \prime }}%
\right) ^{-\alpha }\,\,\,[M_{\odot }\mathrm{pc}^{-3}]\,,
\end{equation}%
with $\alpha =2.0\pm 0.1$ at $R\geq 10^{\prime \prime }$, and $\alpha
=1.4\pm 0.1$ at $R<10^{\prime \prime }$. We have also generalized the
Maximal Likelihood (ML) analysis of the cusp's slope (Alexander 1999) to
take into account the incompleteness corrections. Because the radius of the
extracted NAOS/CONICA field ($\sim $8'') is smaller than the break radius of
10'' indicated by the NTT/SHARP data, an ML analysis of the NAOS/CONICA data
alone cannot be used for a reliable determination of the parameters of the
outer power-law. An analysis of 856 stars in the inner 4'', where the inner
cusp dominates the total counts, indicates an inner power-law cusp with an
exponent $\alpha =1.3\pm 0.1$, in very good agreement with our analysis of
the binned data above. With these parameters the cusp's stellar density is $%
3\times 10^{7}\,M_{\odot }$ pc$^{-3}$ at $R=1^{\prime \prime }$, and $%
7\times 10^{8}\,M_{\odot }$ pc$^{-3}$ at $R=0.1^{\prime \prime }$. An
alternate description of the data in Figure 7 may be a localized,
Plummer-model like, cusp on SgrA$^{\star }$ superposed on a larger-scale,
isothermal cluster, with a distinct break in between (Mouawad et al. 2003).
The present data cannot discriminate between these two possibilities but the
estimated stellar densities are similar. The stellar mass contained in the
cusp is estimated to vary as $1.3\times 10^{4}\,\left[ \frac{R}{arcsec}%
\right] ^{1.63}M_{\odot }$.

The analysis we just presented depends critically on the assumption that the
ratio of number counts to total stellar mass (mostly in fainter, not
directly observed stars) does not vary with radius and environment in the
Galactic Center.\ This assumption is almost certainly violated at some
level, since we argue below that giant late type stars are destroyed and
moderately massive stars are created by mergers of lower mass stars in the
very dense inner region. However, we expect that these effects alter mostly
the counts of the less numerous, brighter stars and not so much the fainter
stars that dominate the counts. At the very least, the radial population
changes make the values of the cusp density estimated above quite uncertain.
Obviously, spectroscopic observations will be required for studying the
properties of the fainter stars as a function of distance from SgrA*.

Keeping this caveat in mind, the observed stellar density distribution is
consistent within the uncertainties with the predictions of theoretical
models for a cluster of stars surrounding a massive central black hole.
These models predict the formation of a power-law cusp. The expected radial
slope of the power-law ranges between $\sim $0.5 and $\sim $2.5, depending
on the cusp's formation scenario and on the importance of inelastic stellar
collisions. Relaxed, single mass stellar cusps have a steep slope of $\alpha
\sim 7/4$ (Bahcall \& Wolf 1976, 1977). Unrelaxed, initially isothermal
clusters around an adiabatically growing hole have a shallower slope of $%
\alpha \sim 3/2$ (Young 1980). In multi-mass, lower density cusps ($\rho
_{\star }<10^{7}\,M_{\odot }$ pc$^{-3}$) the models of Murphy et al. (1991)
also predict a steep slope ($\alpha \sim 7/4$), while higher density cusps ($%
\rho _{\star }\sim 10^{8}\,M_{\odot }$ pc$^{-3}$) have flatter inner slopes
due to the onset of stellar collisions ($\alpha \geq 1/2$). The models also
predict that a large fraction ($\sim $80$\%$) of the cusp stars near the
hole should be bound to it ($\sigma ^{2}=v_{c}^{2}/(1+\alpha )<v_{c}^{2}$, \
where $v_{c}^{2}=GM_{BH}/R,$ Alexander 1999, 2002). Adiabatic solutions
(e.g.Young 1980) that assume that the black hole grows on a time scale that
is short compared to the stellar relaxation time scale, do not apply to the
Galactic center, which is estimated to be relaxed (e.g.Alexander 1999) and
where the growth time scale of the hole is long ($\sim $10 Gyr).

\subsection{K-band luminosity function: old star cluster with an admixture
of young stars}

Figure 9 shows the K-band luminosity function (KLF) for the overall nuclear
cluster ($p<9^{\prime \prime }$). Fig.10 is a color-magnitude plot of the
same region. Figure 11 shows the KLF of the $p\leq 1.5^{\prime \prime }$
cusp region. In both regions we have corrected the counts for incompleteness
with the artificial star technique described in section 2.2, taking into
account the effects of this correction on the error bars. The NAOS/CONICA
data are in excellent agreement with the previous Keck and NTT data at the
brighter magnitudes, and extend the KLF to our completeness limit of K$%
_{s}\sim 18$. The Galactic center KLF thus samples all giants and
supergiants, as well as the main sequence to spectral type A5/F0 (2$\,\
M_{\odot }$). Since the NAOS/CONICA counts are incomplete at the brightest
magnitudes because of saturation effects, we have combined the NAOS/CONICA
counts and the NTT\ counts to a common KLF shown in Figs. 9 and 11.

The overall KLF of the central $p<$9$^{\prime \prime }$ ($0.36$ pc) region
to first order is described by a power law ($\mathrm{d}\log \mathrm{N}/%
\mathrm{dK=\beta }\sim 0.21\pm 0.02$). In the range 14$\leq $K$_{s}\leq $19
the overall KLF is similar to, but somewhat flatter than the KLF of the
Bulge of the Milky Way several degrees from the center ($\beta \sim $0.3:
Alexander and Sternberg 1999, Tiede et al. 1999, Zoccali et al. 2002), and
the KLF on 30pc scales around the center (Figer 2002). A $\beta \sim $0.3
power-law is well matched by the theoretical KLF of old stellar populations,
which reflects the rate of evolution of individual stars along the red giant
and asypmtotic giant branches. The flatter slope of the Galactic center KLF
compared to the Bulge is mostly caused by an excess of the counts at K$\leq
14$, by a factor 1.4 to 2. In agreement with earlier discussions (Lebofsky
\& Rieke 1987,\ Blum et al. 1996b, Davidge et al. 1997), we attribute this
bright-end excess to young, early and late type stars.\ In addition, the $p<$%
9'' KLF shows a prominent excess hump centered at K$_{s}\sim 16$, \ a factor
of 2 above the power law. This hump is prominent in the Bulge as well, and
on 30 pc scales around the Galactic center when all distributions are
adjusted to the same (Galactic center) extinction (A(K)=3.2, Rieke, Rieke \&
Paul 1989, Raab 2002). The excess can be attributed to old and metal rich,
core He-burning horizontal branch (HB)/red clump (RC) stars (Tiede et al.
1995). These stars have characteristic masses of 0.5 to 0.8 $M_{\odot }$.
The red clump stars can also readily be seen in the color magnitude plot.
The amplitude of the HB/RC excess relative to the numbers of giant stars
making up the power law component suggests that the Galactic Bulge on
average has close to solar metallicity (Tiede et al. 1995). In fact, an old (%
$\sim $ 10 Gyr), single age (SSP) model with a bulge metallicity
distribution deduced from the color-magnitude properties of the Bulge gives
a fairly good representation of the hump (dashed curve in Figs. 9 and 11,
from Zoccali et al. 2002). We thus conclude that the hump in the KLF at K$%
_{s}\sim $16 is most likely caused by the presence in the innermost parsec
of such old low mass, and metal rich stars. The overall KLF of the Galactic
center thus is dominated by an old cluster with an admixture of bright young
stars, in excellent agreement with the spectroscopic information collected
over the past decade (see Introduction).

\bigskip

\subsection{Population changes in the cusp}

The KLF of the $p\leq $1.5$^{\prime \prime }$ cusp region (Fig. 11) appears
to be a featureless power-law. It has a similar slope as the integral KLF ($%
\beta =\mathrm{d}\log N/\mathrm{dK}\sim 0.21\pm 0.03$) but the K$_{s}\sim $%
16 HB/RC hump appears to be absent. Compared to the $p\leq $9$^{\prime
\prime }$ and Bulge KLFs this deficit is significant at the 4 to 5 $\sigma $
level. Relative to the $p\leq $9$^{\prime \prime }$ KLF, there may also be
an excess of stars at 13$\leq K_{s}\leq $15 (Fig.11). We conclude that the
cusp probably lacks old low mass, HB/RC stars. The cusp probably also lack
late type giants. The fraction of K$\leq $15 late type stars, as determined
from the narrow-band GEMINI indicator, drops from 50\% at $p\geq 5^{\prime
\prime }$ to 25\% at $p\leq 1.5^{\prime \prime }$(Fig.7). Considering again
the effect of the fore- and background cluster stars, the intrinsic late
type fraction in the cusp region proper must even be lower. Assuming a
simple two-component model with a late type fraction of $\ $50\% in the
outer region, the cusp's content of K$\leq $15 late type stars must
intrinsically be $\leq $15\%. However, because of the small numbers of late
type stars in the very center (7 within 1.5$^{\prime \prime }$) the
statistical significance of this difference naturally is only 3$\sigma $. We
conclude that the content and/or the properties of the stars in the cusp are
different from those in the outer regions. The dense cusp may lack old, low
mass stars. Alternatively and more likely, in the dense environment low mass
stars ascending the red giant branch are stripped of their entire envelopes
by physical collisions or close tidal encounters (see below and Appendix A).
Losing their envelopes well before their helium cores reaches critical mass
for helium ignitions, these stars would evolve directly to become helium
white dwarfs, skipping the HB/RC phase. This could account for the absence
of the HB/RC hump in the KLF and the lower fraction of red giants.

Taking the KLF of the cusp, we can estimate the number of stars as a
function of K that are expected to reside within different radii from SgrA$%
^{\ast }$. We list these numbers in Table 2 for different magnitude (or
mass) limits, for R=1'', 0.1'' and 0.03'', and for two values of $\beta $.
The value of $\beta $=0.21 is the slope determined from the observed KLF for
K$_{s}\leq $ 18. A value of $\beta $=0.35 fits the Bulge KLF (Alexander \&
Sternberg 1999), or a model of an old star cluster (Zoccali et al. 2002,
Fig.11). To normalize to the observations, we use the value of 56 stars
within p$\leq 1"$ and K$_{s}\leq $ 17, as determined from the broken
power-law fit in Fig.7.

The R=0.03'' numbers are interesting since they give an impression of the
number of stars with orbital periods $\leq $2 years and orbital velocities $%
\geq $3000 km/s that may be accessible to future interferometric
observations of the SgrA$^{\ast }$ cusp. For the likely range of the
faint-end KLF slope and for a realistic observational limit of K$_{s}\leq $%
20, Table 2 suggests that one ought to expect one or several such stars to
be present at any given time if they move on circular orbits. The present
data suggest a large fraction of highly eccentric orbits (Sch\"{o}del et al.
2003). In that case the number of fast moving stars would be significantly
larger. In fact two such stars (S2 and S14) have already been observed (Sch%
\"{o}del et al. 2002, 2003, Ghez et al. 2003). The prospects thus are good
for using interferometry to sample stars closer to the central black hole
than presently possible with AO on 10m-class telescopes.

The total number of unobserved, low mass stars (0.5-1 M$_{\odot }$)
extrapolated with the KLF-slope, in combination with the mass model of
section 3.2 above, can also be used to compute the average stellar mass in
the cusp. For $\beta =0.21$ this average mass is 33 M$_{\odot }$ to a
limiting magnitude of K$_{s}=$21 (corresponding to 1 M$_{\odot }$ main
sequence stars), and 13 M$_{\odot }$ to K$_{s}=$23 (corresponding to 0.5 M$%
_{\odot }$ main sequence stars). For $\beta =0.35$ the average masses to the
same K$_{s}$ limits are 8.9 and 1.8 M$_{\odot }$. Assuming a Salpeter mass
function (for lack of a better choice) with power law slope in density $%
\gamma =2.35$ and with lower mass limits of m$_{\min }=$0.5 and 1 M$_{\odot
} $, the average stellar masses are 1.6 and 3.1 M$_{\odot }$, respectively.
For a shallower initial mass function of $\gamma =1.5$, the corresponding
average stellar masses are 7 M$_{\odot }$ (m$_{\min }=0.5$ M$_{\odot }$) and
10 M$_{\odot }$ (m$_{\min }=1$ M$_{\odot }$). KLF slope, number counts and
mass model thus are consistent with each other in either one of two regimes.
If the cusp has a Salpeter IMF, then the faint-end KLF must have a slope
significantly steeper than $\beta =0.21$ and extend to 0.5 M$_{\odot }$, or
lower masses. Alternatively, if the faint-end KLF slope remains near the
value of $\beta $=0.21 characteristic of the K$_{s}\leq $18 data, then the
IMF must be significantly shallower than a Salpeter IMF, thus containing
more high mass stars. The latter explanation would be favored if indeed
stellar mergers are effective, as argued in section 4.4. The difference in
number counts at K$_{s}\sim $19 between a $\beta $=0.35 and a $\beta $=0.21
KLF is about a factor of 2. Ultra-deep H-band NAOS/CONICA observations may
be able to test whether such a steepening of the faint-end KLF does occur.

\bigskip

\subsection{\protect\bigskip Young, massive stars in the central star cluster%
}

There are several pieces of evidence that the cusp contains many young,
massive stars. Moderate resolution, infrared spectroscopy is available for
many bright stars with K$\leq 13$\ and $p\geq 1^{\prime \prime }$ (Genzel et
al. 1996, 2000, Paumard et al. 2001, Ott et al. 2003). The data, as
summarized in Ott et al. (2003), show that 5 of 7 (6 of 11) stars at $p\leq
3^{\prime \prime }$ with K$\leq 11.5$ (K$\leq 12$) are early type, hot
emission line stars (`HeI' stars). The K$_{s}\leq 13$ early type stars at $%
p\geq 1^{\prime \prime }$ must be young. Their near-IR spectral properties
indicate that they are blue supergiants (Of), luminous blue variables (LBVs)
or late type, Wolf-Rayet (WN/C ) stars with ages of 2-7 Myrs and masses of
30 to 120 $M_{\odot }$ (Krabbe et al. 1995, Tamblyn et al. 1996, Najarro et
al. 1997, Ott, Eckart \& Genzel 1999, Paumard et al. 2001, Figer et al.
1998, 1999, Cotera et al. 1999, Figer 2002). There is a prominent
concentration of these stars in the IRS16 and IRS13 complexes. Ott et al.
(1999) found that the bright HeI emission line star IRS16 SW is an eclipsing
binary of mass $\sim 100\,M_{\odot }$. The alternative possibility that the
hot emission line stars are rejuvenated products of mergers from lower mass,
old stars can be excluded for stars outside the central few tenths of an
arcsecond (see Appendix A). The stellar density several arcseconds or more
away from SgrA$^{\star }$ is definitely too low to account for mergers to
produce stars more massive than 10 to 20 $M_{\odot }$ (Genzel et al. 1994,
Lee 1994). Furthermore, the properties of the bright emission line stars are
essentially identical to the massive stars in the Quintuplet and Arches
clusters, 30 to 50 pc away from SgrA$^{\star }$, in much lower density
environments (Figer et al. 1998, 1999, Cotera et al. 1999, Figer 2002)..

Within the `SgrA$^{\star }$ cluster' ($p\leq $0.5$^{\prime \prime }$) Gezari
et al. (2002) and Ghez et al. (2003) have recently reported adaptive optics
assisted spectroscopy of several stars with the Keck telescope. S2 exhibits
HI Br$\gamma $ absorption and clearly is a hot star with an equivalent
main-sequence identification of O8/B0. S1 and the fainter star S0-16 do not
exhibit CO absorption features, and thus are very likely early type stars,
with equivalent main sequence identifications of B0 (S1) and B5 (S0-16)
(Genzel et al. 1997, Eckart et al. 1999, Figer et al. 2000, Gezari et al.
2002). Based on narrow-band, speckle-spectro-photometry Genzel et al. (1997)
concluded that S8 and S11 are early type stars as well. The `S'-stars near
SgrA$^{\star }$ represent the possible K$_{s}\sim $14 excess found in the
KLF.

\bigskip

\subsection{Two kinematic components of young stars}

We next use the narrow-band colors and and proper motions of the Ott et al.
(2003) sample to demonstrate that the somewhat fainter (K$_{s}\leq $ 15),
spectro-photometrically identified early type stars in the central few
arcsecs must also be young and can be separated into two, separate kinematic
components. Consider the normalized angular momentum along the line of
sight, $J_{z}/J_{z}(\max )$, which we define as 
\begin{equation}
J_{z}/J_{z}(\max )=(xv_{y}-yv_{x})/pv_{p}\,,
\end{equation}%
where $v_{x},v_{y}$ and $v_{p}$ are the R.A.-, Dec.- and total proper motion
velocities of a star at (x,y) on the sky and at projected radius $p$. $%
J_{z}/J_{z}(\max )$ is $\sim -1$, $\sim 0$ and $\sim +1$, depending on
whether the stellar orbit projected on the sky is mainly counter-clockwise
tangential, radial or clockwise tangential with respect to the projected
radius vector from the star to SgrA$^{\star }$. Note that stars moving
tangentially in projection also must move tangentially in three dimensions.
Stars moving radially in projection, however, may move radially in 3D, or
tangentially, or a mixture of the two, depending on their location along the
line of sight. Figure 12 shows the distribution of $J_{z}/J_{z}(\max )$ as a
function of \ $p$ for late and early type stars, as identified from the
narrow-band CO-index (Fig. 6). Figure 13 is a plot of the spatial
distribution on the sky of early and late type stars with clockwise,
counter-clockwise and radial orbits. For convenience, we divided here the
normalized angular momentum range into 3 bins: clockwise tangential ($%
J_{z}/J_{z}(\max )\geq 0.6$), counter-clockwise tangential ($%
J_{z}/J_{z}(\max )\leq -0.6$) and radial (-0.3$\leq J_{z}/J_{z}(\max )\leq $%
0.3). The isotropic late type stars serve as our comparison sample. They
exhibit a random distribution of line-of-sight angular momenta (left panel
of Fig. 12, right panel of Fig. 13). 21($\pm 9$)\% (22($\pm 3$)\%) and 36( $%
\pm 11$)\% (31($\pm 4$) \%) of the late type stars within $3^{\prime \prime
} $(10$^{\prime \prime }$) are on clockwise and counter-clockwise orbits. In
contrast the early type stars show a preponderance of tangential and a lack
of radial orbits (Fig. 12 right panel). Within the central $3^{\prime \prime
}$ 51($\pm 10$) \% \ and 18($\pm 6$) \% of the early type stars are on
clockwise, and counter-clockwise, tangential orbits. Between 3 and 10''
there are about equal numbers of clockwise and counter-clockwise tangential
motions but still a lack of radial orbits. The preponderance of clockwise,
tangential velocities near SgrA$^{\star }$ is also graphically apparent in
the overdensity of filled circles within 3'' from SgrA$^{\ast }$ in the left
inset of Fig.13. In marked contrast to the random pattern of the late type
stars broken up into the same three groups (Fig. 13, right panel), early
type stars with clockwise, tangential orbits dominate within a few
arcseconds of SgrA$^{\star }$, and bunch up in the IRS16 complex E/SE of SgrA%
$^{\star }$. This means that most of the early type stars in the central
10'' are not relaxed, and belong to one of two kinematic components with
opposite, line-of-sight angular momenta.

\bigskip

The line-of sight velocities also confirm that the late type stars are
relaxed but early type stars are not. Apart from an average blue-shift in
the central 10'', the late type stars show a random distribution of
line-of-sight velocities (McGinn et al. 1989, Sellgren et al. 1990, Genzel
et al.1996, Haller et al. 1996). In contrast, early type stars north of SgrA$%
^{\star }$ are almost all blue-shifted, while stars south of SgrA$^{\star }$
are almost all red-shifted Genzel et al. 1996, 2000, Paumard et al. 2001).
This clearcut pattern is shown in the right inset of Fig.14 \ and is
indicative of coherent rotation, with a direction opposite (counter) to that
of the overall Galactic rotation.

\bigskip

\subsection{\protect\bigskip Two coeval, rotating disks of young stars}

This global rotation pattern has previously been discussed by Genzel et al.
(1996, 2000)\ and Paumard et al. (2001). Genzel et al.(2000) found that the
3D-velocity field could be matched by a rotating disk at inclination $\sim $%
140$^{o}$ (or 40$^{o}$, depending on the definition of the angles) but noted
that the fit was poor, \ and that ' a better description .... probably is a
dynamically hot and geometrically thick, rotating torus ....'. In the
following we use the new data set of Ott et al. (2003) for a better
determination of the disk parameters and for an investigation whether the
two kinematic components of early type stars discussed in the last section
can both be fit by rotating disks. For this purpose, we followed Levin and
Boloborodov (2003) and first analyzed all 14 clockwise (J$_{z}/$J$_{z}$(max)$%
\geq $0) early type stars with three velocities, as well as all 12
counter-clockwise (J$_{z}/$J$_{z}$(max)$\leq $0)\ early type stars with all
three velocities measured. We explored whether there exists a disk of
inclination $i$ and projected major axis position angle (line of nodes) on
the sky $\phi _{o}$ that fits the velocity field of all stars. We searched
for a vector $\overrightarrow{P}$($i,\phi _{o}$) (perpendicular to the disk)
for which the observed velocities of the stars along $\overrightarrow{P}$
are minimized. For the 14 clockwise stars we find that the best fitting
plane ($i=-120\pm 7^{o}$ and $\phi _{o}=-60\pm 15^{o}$) fits the data with $%
\chi _{red}^{2}\sim $ 4 (Fig.15, left inset). Most of the deviation from a
common plane is caused by one star located 9.8$"$ SW of SgrA$^{\ast }$%
(AFNW). Our proper motions at such large distances strongly depend on the
assumed values for the second order distortion terms in our astrometric
solution. Since these terms are not as well constrained by the SiO maser
stars in the field as the first order terms, there may be large systematic
uncertainties for proper motions this far from the center. Taking only stars
with separations 7'' or smaller from SgrA*(filled circles in Fig.15, left
inset) reduces the number of stars to 12, and $\chi _{red}^{2}$ to 2. The
clockwise early type stars with three velocities thus can be described very
well by a thin, rotating disk (at average distance of 2-4'' from the
center). The derived disk orientation agrees with that obtained by Levin and
Boloborodov (2003) who used the earlier Genzel et al. (2000) proper
motion/radial velocity set. A position angle of -40 to -60$^{o}$ is also
consistent with the spatial elongation of the\ stars in the x/y plane
(Fig.14, left inset, with the exception of AF and AFNW 9-10'' SW). For the
12 counter-clockwise stars the best fitting plane ($i=-40\pm 15^{o}$ and $%
\phi _{o}=160\pm 15^{o}$) fits the data to $\chi _{red}^{2}=$ 5 (Fig.15,
right inset). Removing the two stars more than 7'' away from SgrA* (open
squares) reduces $\chi _{red}^{2}$ somewhat to 4.8. Again, a low inclination
disk is consistent with the spatial distribution of the counter-clockwise
stars in Fig.14 (left inset). A rotating thin disk model thus also fits the
counter-clockwise stars but the deviations from the model are somewhat
larger than for the clockwise stars. The characteristic radius of the
counter-clockwise stars is 4-7''.

\bigskip

In the next step we explored whether also the many other clockwise and
counter-clockwise stars with proper motions (but without 3 velocity
components) can be fit by the same disk parameters. From coordinates and
proper motions one can construct two angles, $\alpha (x,y)$ and $\beta
(v_{x},v_{y})$ ($\ \tan \alpha =x/y,\tan \beta =v_{x}/v_{y}$), which are
independent of the radius in the disk but are functions of $i$ and $\phi
_{o} $. The thick curve in Fig.16 denotes the loci of stars in a disk with
parameters derived above for the clockwise stars, while the other curves
denote model disks with parameters within the uncertainties of the best fit.
The filled and open circles denote the clockwise tangential stars (J$_{z}/$J$%
_{z}$(max)$\geq $0.6) for p$\leq $5'' and K$\leq $14 and 15, respectively.
Clearly these fainter stars fit the input disk model (of the 14 stars with 3
velocities) very well. Likewise the counter-clockwise tangential stars fit
the input disk model of the 12 stars with 3 velocities. In contrast late
type stars selected in the same manner (J$_{z}/$J$_{z}$(max)$\geq $0.6, K$%
\leq $14, 15) do not fit a rotating thin disk model. They exhibit a large
scatter around the best fit disk model (right inset of Fig.16). A similar
check can also be done with the counter-clockwise stars. However, because of
the small inclination, the data are not as well constrained as in the
clockwise subset.

We conclude that the dynamics of most early type stars can be well fit by
two rotating and geometrically fairly thin disks. Levin and Boloborodov
(2003) conclude from their analysis that the (clockwise) disk may be
geometrically infinitely thin. Our own analysis suggests that a finite
thickness, or moderate warping of the disks probably is probably still
consistent with the data, especially for the counter-clockwise stars. Figure
17 summarizes the orientation of the two derived disks. The two disks are at
large angles relative to each other, effectively counter-rotate, but do
share a common projection of its rotation vector that is opposite to that of
Galactic rotation (Fig.14, right inset). The clockwise disk is more compact,
with a characteristic radius of 2-4'', while the conter-clockwise disk has a
characteristic radius of 4-7''.

\bigskip

The stellar content of the two disks is essentially identical. Of the 14
clockwise stars with spectroscopic identifications in Ott et al. (2003),
there are 4 Of/LBV, 5 WNL (7-9), 1 WNE (5/6) and 4 WCL (8/9) stars. Of the
12 spectroscopic counter-clockwise stars, there are 2 Of/LBV, 3 WNL and 6
WCL and 1 WCE (5/6). The agreement of the different subtypes of blue
supergiants and Wolf-Rayet stars is remarkable. The lifetimes of these
different subtypes is no more than a few hundred thousand years and the WC
phase occurs just before explosion as supernova (Maeder \& Meynet 1994).
Hence the two disks must have formed at the same time, within less than 1
Myr.

\bigskip

The early type (`S'-) stars in the `SgrA$^{\star }$ cluster' ($p\leq
0.5^{\prime \prime }$) show a different angular momentum distribution
(Fig.12, right panel). Keeping in mind that a few of these stars may only be
projected to lie in the central arcsecond, the `SgrA$^{\star }$ cluster'
stars do not show an excess of tangential orbits. On the contrary, a more
detailed analysis shows an excess of radial orbits (Genzel et al. 2000, Sch%
\"{o}del et al. 2003). 54 ($\pm 14$) \% of the 35 observed stars in the `SgrA%
$^{\star }$ cluster' are on radial orbits. These stars are probably observed
near the apoapse of their orbit (roughly twice the semi-major axis), where
they spend most of their time. It is therefore likely that the radial stars
are tightly bound to the black hole, with semi-major axes of order 20 mpc.
This statistical inference, together with the direct derivation of
semi-major axes of $<$10 mpc for the 6 stars in the inner cusp whose orbits
were solved (Schoedel et al 2003), suggests that the Sgr A* cluster as a
whole is tightly bound to the black hole. The orbital planes of those stars,
as far as they are constrained by the current data, are significantly
different from the two disk planes of the early type stars at $p\geq $1''.
For instance S2 has a clockwise angular momentum, but its orbital plane is 70%
$^{o}$ off the plane of the clockwise early type stars discussed above.
Levin and Beloborodov (2003) have commented that such large offsets may be
explained by relativistic Lense-Thirring precession of a star originally in
the same plane as the stars further out.

In summary, we find compelling evidence from different arguments - KLF,
spectra/narrow-band photometry, dynamics - that the stellar cusp centered on
the massive black hole candidate SgrA$^{\star }$ does not primarily consist
of old, low mass stars with properties similar to those found in the
Galactic Bulge. Late type giants and horizontal branch, red clump stars seem
to be less frequent. The cusp contains a suprising number of unrelaxed and
apparently young, massive stars, including the very short lived HeI emission
line stars with masses 30-100 M$_{\odot }$ that reside in two rotating
disks. This result is not consistent with theoretical expectations for a
steady-state cusp, which predict that such a cusp should be dynamically
relaxed\ (Bahcall \& Wolf 1976, 1977, Murphy et al. 1991). The properties of
the nuclear region are not consistent with static equilibrium. Episodic star
formation and stellar transformations are required.

\bigskip

\subsection{Mid-IR excess of S2: probing the accretion flow onto SgrA$%
^{\star }$?}

Table 2 lists the H/K$_{s}$/L$^{\prime }$ photometry of S2/SgrA$^{\ast }$
and 11 other stars in the central SgrA$^{\star }$ cluster as derived from
the new NAOS/CONICA photometry. At H and K$_{s}$ the peak emission of S2/SgrA%
$^{\ast }$ is definitely centered on S2 (and off SgrA$^{\star }$), and we
can only deduce upper limits to the magnitudes of SgrA$^{\star }$ (Figure 3,
Table 2). The H-K$_{s}$ color of S2 (1.3$\pm $0.1 mag) compares well to the
average value of the 11 other, H$<\!17$ stars within 0.7$^{\prime \prime }$
of SgrA$^{\star }$($\langle $H-K$_{s}\rangle $=1.42 $\pm $0.03). This color
is consistent with that of blue stars (\{H-K$_{s}$\}$_{(0)}$=-0.04) behind
the standard Galactic center K-band extinction of \ $\sim 3$ mag (Rieke,
Rieke \& Paul 1989), and thus is consistent with their being early type
stars, as discussed above. However, the K$_{s}$-L$^{\prime }$ color of
S2+SgrA$^{\ast }$ (2.9$\pm $0.1 mag) is 0.65 mag redder than the same eleven
stars ($\langle $K$_{s}$-L$^{\prime }\rangle =2.25\pm 0.15$). S2+SgrA$%
^{\star }$ thus has a significant L$^{\prime }$ excess relative to stars
with the same H-K colors (see also Clenet et al. 2003). Extrapolating from
the average color of the nearby stars, the excess flux density corresponds
to 4.9 mJy. Correcting for a K-band extinction of 3.2 mag, or an L$^{\prime
} $ extinction of 2.1 mag (Raab 2002: A(H)=5.1, A(K$_{s}$)=3.2 and A(L$%
^{\prime }$)=2.1), the dereddened L$^{\prime }$-band excess flux density of
S2+SgrA$^{\star }$ is about 33 mJy. Gezari et al. (2002) and Ghez et al.
(2003) find that the K-band flux density and spectrum of S2 is consistent
with a O8/O9 main sequence star, which would have a mass of 15-20 $M_{\odot
} $, a bolometric luminosity of $10^{5}\,L_{\odot }$ and a temperature of
35,000 K. Such a main sequence star would not be expected to have a very
dense, dusty wind which could account for the observed L$^{\prime }$
emission. What then causes the L$^{\prime }$ excess of S2?

\bigskip

\bigskip One obvious possibility is that the excess is due to emission from
SgrA$^{\ast }$ itself in which case we have the first detection of the
mid-infrared emission from the accretion region in the immediate vicinity of
the black hole. This explanation may be supported by the spatial
distribution of the L$^{\prime }$ emission. The peak of the L$^{\prime }$
emission is closer to SgrA$^{\ast }$ and the L$^{\prime }$ emission of
S2/SgrA$^{\ast }$ appears to be spatially extended (lower right inset of
Fig.3). \ The FWHM of S2+SgrA$^{\ast }$ on the L$^{\prime }$ Lucy-map
(Fig.3) is 95$x$70 mas (EW $x$ NS), compared to 70x70 mas for unresolved
stars. Current models of the accretion flow predict that the infrared
emission predominantly comes from the tail of the radio/submm synchrotron
emission (Melia \& Falcke 2001, Narayan et al. 1995, Falcke \& Biermann
1999, Markoff et al. 2001, Liu \& Melia 2002, Hawley \& Balbus 2002). The
infrared spectral energy distribution is inverted and the L$^{\prime }$-band
data provide valuable constraints on the iinnermost parts of the accretion
zone. Our deduced L$^{^{\prime }}$ excess flux density is significantly
larger than the values predicted by the various theoretical, `quiescent
state' models (Markoff et al. 2001, Liu \& Melia 2002). The model that comes
closest (within a factor of a few) is the jet-disk model of Markoff et al.
(2001). Enhanced mid-IR emission at about the observed level is
theoretically predicted to occur during flares, such as observed at X-rays
(Baganoff et al. 2001).

Alternatively the mid-IR emission may come from dust in the accretion flow
onto the black hole that is heated by the hot and luminous O star S2. We
show in Appendix B that dust in the accretion flow onto SgrA$^{\ast }$ can
be heated to sufficently high temperature by the ultra-violet radiation from
S2 and can account for the L$^{^{\prime }}$ excess if 
\begin{equation}
\left[ \dot{M}f_{\mathrm{dust}}\right] _{R=0.002\mathrm{pc}}\sim 2\times
10^{-7}\,\ [M_{\odot }\,\mathrm{yr}^{-1}]\,.
\end{equation}

Here $\dot{M}$ is the mass accretion rate onto the black hole and $f_{%
\mathrm{dust}}$ is the abundance of dust in the flow, relative to that in
the interstellar medium. This mass accretion rate is in good agreement with
a recent estimate (10$^{-7}$ $M_{\odot }$ yr$^{-1}$) based on the $\sim $%
10\% linear polarization of SgrA$^{\star }$ in the mm-range (Bower et al.
2002). The estimated mass accretion rate also agrees with recent theoretical
models of the SgrA$^{\star }$ spectral energy distribution, including its
X-ray emission (Liu \& Melia 2002, Markoff et al. 2001), but is two to three
orders of magnitude lower than earlier models based on advection flows
(ADAFs: Narayan, Yi \& Mahdevan 1995) or Bondi flows (Melia 1992).

\bigskip

A definite test whether the observed L'-band excess is caused by S2 heating
the accretion flow or by (variable) emission from the black hole flow itself
will come from multi-epoch observations that we have begun on the VLT in the
spring of 2003. The 'UV-heating' model predicts that the excess emission
should continue to be centered on S2 but should become dimmer as the star
moves away from Sgr A. The SgrA$^{\ast }$ model in turn predicts that the
mid-IR excess emission should be centered on the radio source, probably be
polarized and strongly vary with time. First observations taken in March
2003 appear to favor that the source of the excess emission is SgrA$^{\ast }$
itself.

\bigskip

\section{\protect\bigskip Discussion: The origin of the early type stars in
the cusp}

In this section we will discuss how the massive and apparently young stars
discussed in the last section may have come to reside in the dense central
environment around the central massive black hole. The key questions are
whether they can have formed in situ, or whether otherwise they might have
migrated there from further out. We will discuss different possibilities
that have been proposed and discussed in the literature during the last
decade. We believe that our new dynamical data make a good case that a cloud
collision mechanism is the most likely possibility for the formation of
massive stars at 1-10'' from the center, while a collision-merger model of
smaller stars may be the most likely model for the formation of the
'S'-stars in the innermost region around the black hole.

\bigskip

First, the blue supergiants and Wolf-Rayet stars 1-10'' from the center, and
even more clearly the 'S'-stars in the SgrA* cluster cannot have migrated
through two-body processes to their present location from large radii. The
relaxation time of stars of mass $m$ is given by (e.g.Alexander 2002) 
\begin{equation}
t_{r}(m)=10^{8.2}M_{3}^{3/2}R_{10}^{-0.13}m_{10}^{-1}(\ln (N_{\star
}/12))^{-1}\,\,\,\,[yr]\,,
\end{equation}%
where $M_{3}$ is the mass enclosed within radius $R_{10}$ (in units of 10$%
^{\prime \prime }$ or 0.38 pc) in units of $3\times 10^{6}$ $M_{\odot }$, $%
m_{10}$ is the stellar mass in units of 10 $M_{\odot }$ and $N_{\star }$ is
the number of stars. In the central 10$^{\prime \prime }$ the relaxation
time is approximately independent of radius (for a cusp of radial exponent $%
\alpha $=1.4). This needs to be compared to the (main sequence) stellar
lifetime (Genzel et al. 1994, Cox 2000), 
\begin{equation}
t_{ms}(m)=\left\{ 
\begin{array}{ll}
10^{7.3}m_{10}^{-2.5}\,\,\,[yr] & \mbox{for $m_{10}\leq0.9$} \\ 
10^{7.2}m_{10}^{-1.5}\,\,\,[yr] & \mbox{for $m_{10}>0.9$}%
\end{array}%
\right. \,.
\end{equation}%
Clearly, stars more massive than about 2.5 $M_{\odot }$ cannot have formed
outside the central region and then mass-segregated through two-body
relaxation to the central few arcseconds.

\bigskip

\subsection{Current star formation in the mini-spiral?}

One possibility is that massive stars form continuously outside the dense
central region and rapidly pass through the center on highly elliptical
orbits. For instance, the northern arm of the mini-spiral appears to be on
such a trajectory. The dusty sources associated with the gas might be
candidates for young stars formed in and still embedded in their parental
gas clouds. Becklin et al. (1978) were the first to propose that compact,
mid-IR bright sources in the mini-spiral, such as IRS1, 2/13, 5 and 10 are
compact HII regions surrounding intrinsically luminous (and hot) stars.
Gezari et al. (1994) showed that these sources and also IRS21 are local dust
temperature peaks in the mini-spiral and, therefore, must contain local
luminosity sources. Based on the dust-excess, featureless near-IR spectra of
the `mini-spiral sources', Krabbe et al. (1995) and Ott et al. (1999)
proposed that they might be young, recently formed stars embedded in the
mini-spiral. IRS1, 5, 10 and 21 are, however, offset from the emission and
temperature peaks of the diffuse dust in the mini-spiral ridge (Cotera et
al. 1999). Tanner et al. (2002) carried out detailed Keck observations of
the near- and mid-IR emission of IRS21 and proposed that this source is a
luminous post-main sequence (Wolf Rayet) star. In the model of Tanner et al.
IRS21 is just fortuitously near the gas/dust of the mini-spiral. Its
extended dust emission is a bow shock created by the motion of the star into
the mini-spiral. Comparison of the stellar proper motions of these
mini-spiral sources with the proper motions of the gas derived from VLA
images (Yusef-Zadeh et al. 1998, Zhao \& Goss 1999) can distinguish between
the `embedded young star' and `older stars moving through gas/dust'
hypotheses. The results of this comparison are shown in Fig. 18.

The left panel of Fig. 18 shows a grey-scale 1.3cm radio image of the
northern/eastern arms and the bar (Zhao \& Goss 1998), along with some of
the gas proper motion vectors deduced by Yusef-Zadeh et al. (1998). The bulk
of the ionized gas in the northern arm, mini-cavity and bar is streaming
south/south-west in a counter-clockwise pattern on the sky, and in the
general sense of Galactic rotation (Lacy et al. 1991, Yusef-Zadeh et al.
1998). The gas streamers are in a plane, which is inclined $\sim $65$^{o}$
with respect to the line of sight (Lacy et al. 1991, Vollmer \& Duschl
2000). The right panel shows the proper motion vectors of those stars in the
Ott et al. sample that are not late type stars, are dusty, i.e. have a
significant L$^{\prime }$ excess (K-L$^{\prime }$=2), and have K$<15$. These
proper motions clearly demonstrate that most of the dusty mini-spiral L$%
^{\prime }$ stars do not share the motion of the mini-spiral gas. As
proposed by Tanner et al. (2002), they just happen to move through and
interact with the mini-spiral gas/dust. IRS10\ (W), several sources in the
northern arm (NA), IRS1, IRS21, the IRS16 sources and sources in/near the
bar (IRS 29/3/34) have proper motions that are inconsistent with the gas
streaming vectors. In particular, the proper motion of IRS21 is fully
consistent with the Tanner et al. (2002) bow shock model. One of the
northern arm sources moving eastward into the northern arm in fact is
associated with a bow-shock structure on the L$^{\prime }$ map (source NA
moving eastward in the right panel). In addition the spectrosopic data of
Ott et al. (2003) indicate that several of the dusty sources indeed are late
type, carbon-rich Wolf-Rayet (WC) stars. We thus conclude\ that there is
little evidence for current star formation in the mini-spiral gas streamers.
The only possible case for dusty stars moving along with the gas of the
mini-spiral is the IRS13 complex. At the northern tip of the IRS13 complex
is a group of compact L$^{\prime }$ sources with a very large K-L$^{\prime }$
excess, which may be candidates for deeply embedded and very young stars
embedded in the bar (Eckart et al. 2003). The IRS13 complex may thus be a
plausible candidate region for recent star formation in the gas streamers.
It is not clear, however, whether IRS13/2 is actually part of the large
scale, northern arm structure, or whether it is a separate dynamical feature
in the bar (Vollmer \& Duschl 2000).

If massive stars are forming frequently in dense gas streamers when outside
the central parsec and then rapidly move through the central region, one
would expect $\sim $100 times as many massive stars outside the central
region as in the central parsec. Only a few such stars have so far been
found in the central 10 pc, although complete surveys have been shallow
(Figer 1995) and deeper searches were confined to known compact HII regions
(Cotera et al. 1999). In the recent deeper spectroscopic work by Ott et al.
(2003) no new emission line stars have been detected outside the known
'starburst region' of diameter $\sim $25$"$. With the possible exception of
IRS13, a model in which massive stars are formed outside the dense central
region and just pass through the center on highly elliptical orbits does not
seem to be supported by the observations. Most of the bright dusty sources
associated with the mini-spiral are not embedded `protostars'.

\subsection{Infall of young, massive star clusters}

To overcome the long time scales implied by two-body relaxation, Gerhard
(2001) proposed that a young star cluster spiraled into the nucleus through
dynamical friction, and made it to the central parsec before it was tidally
disrupted there. Portegies Zwart, McMillan \& Gerhard (2003), have shown by
more detailed numerical simulations that the Gerhard scenario is feasible.
It requires that such a cluster has to be very massive ($\gg $10$^{4}$ $%
M_{\odot }$) and dense (and thus compact: 0.2-0.4 pc) to spiral into the
center within the lifetime of its O-stars (a few Myrs), core collapse and
arrive there before dissolving completely. The dynamical friction time
scales inversely with the cluster mass so that a 10$^{5}$ -10$^{6}$ $%
M_{\odot }$ cluster can spiral into the center from parsec-scale distances
in a few Myrs. The two known (non-nuclear) young star clusters in the
central 30 pc, the Quintuplet and Arches clusters, have a diameter of $\sim
0.4$ pc and a mass of a few $10^{4}\,M_{\odot }$ (extrapolating the observed
stellar content to 1 $M_{\odot }$; Figer 2002). The N-body simulations by
Portegies Zwart et al. (2003) show that a cluster of mass 6x10$^{4}$ $%
M_{\odot }$ and diameter $<0.3$ pc could make it into the central parsec
within the available time from an initial radius of 4--5 pc, but not from
further out. The in-spiraling cluster maintains its orbital angular momentum
direction and the tidally disrupted final remnant of such a cluster
resembles a thin rotating disk, similar to what we observe. However, the
final radius reached by the dissolving cluster in the simulation by
Portegies Zwart et al. (2003) is 1-2 pc, significantly larger than the
characteristic radius of either one of the two rotating disks of young stars
(0.1-0.4 pc). Further, there is presently no observational evidence for such
dense ($>10^{6}$ cm$^{-3}$), compact and massive molecular concentrations
within a few parsec of the nucleus (Mezger et al. 1996). Several $>10^{5}$ $%
M_{\odot }$ molecular clouds are located at distances $>10$ pc from the
center (Mezger et al. 1996). At that radius the tidal requirements are much
reduced, although a cluster that formed efficiently from such a cloud must
still have been very compact to rapidly core collapse und thus make it
intact into the central region. Finally the fact that we see two such disks,
with different angular momenta, requires that two massive clusters fell in
at about the same time (and none later). This is statistically unlikely. We
conclude that the 'infalling cluster' scenario, while attractive, fails to
match the specific properties of the two rotating disks of young stars in
the Galactic center.

\subsection{In situ formation from dense gas}

The third possibility is that stars formed in situ from cloud collapse,
following the infall of (a) dense gas cloud(s). Several molecular and
ionized gas clouds are detected in the central parsec, several with large
radial velocities, and a few even with the correct (and anti-Galactic)
angular momentum (Jackson et al. 1993, Lacy et al. 1991, Genzel et al.
1996). The mass in the presently observed circum-nuclear molecular material
is several 10$^{4}$ M$_{\odot }$. \ This is amply sufficient to account for
the massive star content in the 2-7 Myr starburst component 1-10'' from the
center (one to a few thousand solar masses, Krabbe et al. 1995). It may also
be sufficient to account for formation of lower mass stars with a
Salpeter-like, mass function if the star formation efficiency is
sufficiently high. For self-gravity in interstellar clouds to overwhelm the
tidal shear and create $10^{3-4}\,M_{\odot }$ of stars in the 2-7 Myr
starburst component, however, requires gas densities of 5 to 10$\times
10^{9} $ hydrogen atoms cm$^{-3}$. This is several thousand times greater
than the density of gas clumps currently found in the atomic and molecular
gas clouds in the central 1-2 pc ($n(\mathrm{H,H}_{2})\leq 10^{6}$ cm$^{-3}$%
, Jackson et al. 1993, Genzel et al. 1994). For the `S'-stars bound to the
black hole within the central arcsecond the required gas densities are even
more extreme ($\geq $ 10$^{13.5}$ cm$^{-3}$). Hence gas clumps naturally
occuring in known circum-nuclear gas would have to be compressed very
substantially before they can collapse to form stars. A cloud-cloud
collision and shocks resulting from it, as well as the action on that gas by
stellar winds in the central region may be possible agents for achieving
such compressions (Morris 1993). Some evidence for large compression and the
action of winds come from high resolution radio maps (Yusef-Zadeh et al.
1998) and the new NAOS/CONICA L$^{\prime }$ images (Fig. 1, 18, Cl\'{e}net
et al. 2003). The L$^{\prime }$ image exhibits sharp (diameter $\leq
0.1^{\prime \prime }$) filaments, plausibly caused by the interaction of the
mini-spiral streamers with the outflows and winds from the central few
arcseconds. \bigskip

There are no published simulations of such a cloud collision near a black
hole but it is clear qualitatively that the cooling, post-shock material
would have lost a lot of its original angular momentum and settle in a disk
circulating around the hole. Subsequently the disk could loose additional
angular momentum through magnetic friction (Hawley \& Balbus 2001) or
interaction with the stellar winds in the central parsec. Through that
angular momentum transport\ the gas disk could then slowly contract and
eventually become dense enough in its inner parts to form stars (e.g. Levin
and Boloborodov 2003, Nayakshin et al. 2003).

\bigskip

We cannot quantitatively evaluate this scenario at the present time but
conjecture that it has two natural and critical features matching our new
observations. First a cloud collision is a sudden compression that may
trigger a burst of star formation. Second the two colliding clouds would
plausibly move in rather different directions and would be compressed at the
same time, thus leading to coeval star formation in the post-shock gas of
both clouds. If the splashing, shocked gas retains some memory of the
original cloud momenta, the gas might settle in two, counter-rotating disks.
We therefore think that the 'colliding clouds' model is qualitatively very
attractive for explaining the two counter-rotating and coeval massive star
disks 1-10'' from SgrA$^{\ast }$. Obviously a realistic hydrodynamical
simulation of such a cloud collision in the gravitational potential of a
massive black hole should be carried out for testing the notion that a cloud
collision can lead to two separate rotating disks. If that turns out to not
be the case, one would have to resort to a model in which two clouds
happened to fall into the center at the same time, and then gradually lost
angular momentum and contracted until they could form stars. This scenario
suffers from the same problem of low statistical probability as for the two
in-spiraling star clusters.

\subsection{The `S'-stars near SgrA$^{\star }$: super-blue stragglers?}

Another possibility is that the massive stars are continuously formed
through mergers of lower mass stars. In this case the massive early type
stars may be old stars that have been rejuvenated in the very dense
environment of the cusp (Morris 1993, Genzel et al. 1994, Lee 1994,
Alexander 2002). The probability for massive stars to be formed in this way
can be qualitatively assessed by comparing the collision time scale between
stars of different masses with their lifetime. Since the collisional time
scale increases but the stellar lifetime decreases with mass, the maximum
stellar mass formed in a merger tree occurs roughly when the two time scales
are equal. As shown analytically in Appendix A, \ 10 $M_{\odot }$ stars can
plausibly be formed in this way in the innermost cusp, within a few tenths
of an arcsecond of SgrA$^{\star }$, if the merger efficiency in high
velocity collisions is high and if angular momentum of the rapidly rotating
merger is efficiently removed. The `S'-stars in the `SgrA$^{\star }$
cluster' may thus be collisional products. They may be 'super-blue
stragglers'. similar to but much more massive than those found in globular
clusters (Bailyn 1995). However, the more massive `HeI' emission line stars
(30-100 $M_{\odot }$) 1-12$^{\prime \prime }$ from SgrA$^{\star }$ clearly
cannot have been formed this way, in agreement with the more detailed
Fokker-Planck calculations of Lee (1994).

Another predicted effect of stellar collisions in the cusp is the
destruction of late type giants (Alexander 1999, 2002). We show in Appendix
A that within the central $\sim $1$^{\prime \prime }$ near-central
collisions occur between late type giants and solar mass stars within a
giant's lifetime. If such collisions permanently destroy the giant's
envelope, \ one would expect that the density of late type giants decreases
in the central few arcsecs. We have shown above (Figs. 7) that the fraction
of K$_{s}\leq 15$ giants in fact appears lower within the central $2^{\prime
\prime }$, which is in agreement with the collider model and the earlier
conclusions of Alexander (1999).

\bigskip

\section{\protect\bigskip Conclusions}

We have presented a new analysis of the properties of the star cluster in
the central 10'' of our Milky Way, \ based on deep, near-IR adaptive optics
imaging with NAOS/CONICA on the ESO VLT, and a new data base of stellar
proper motions. Our main results can be summarized as follows.

\begin{enumerate}
\item The faint stars form a power-law cusp centered on the massive black
hole SgrA$^{\ast }$. Our observations resolve the long-standing paradox of
why the infrared light distribution peaks away from the position of SgrA$%
^{\ast }$. The stellar density in the central arcsecond exceeds 3$x$10$^{7}$M%
$_{\odot }$pc$^{-3}$. Stellar collisions are expected to be frequent.

\item The shape of the K-band luminosity function of the central par$\sec $
is consistent with the properties of an old, metal rich star cluster, with
an admixture of massive young stars.

\item K-band luminosity function, faint number counts and late type star
fraction in the central few arcseconds suggest that stellar content and
properties change significantly in the dense innermost cusp. To the K$\leq $%
15 magnitude limit reached by spectra and proper motions, the cusp is
dominated by massive, early type stars.

\item The K$\leq $15 early type stars, similar to the brightest 'HeI
emission line stars' studied over the past decade, are dynamically unrelaxed
and thus must be young. Most of the early type stars in the central 10
arcseconds appear to reside in two fairly thin, rotating disks. These disks
orbit the black hole, are inclined at large angles and counter-rotate with
respect to each other. The two star disks have essentially the same stellar
content and thus must have formed coevally, within less than 1 Myr.

\item Of the various proposals how the massive stars have made it into the
central 1-10'' our new observations, and in particular the pattern of two
coeval rotating disks of massive stars, now appear to favor in situ
formation following the collision of gas clouds. We propose that two dense
and moderately massive (a few 10$^{3}$ to 10$^{4}$ M$_{\odot }$)
interstellar clouds fell into the central parsec and collided there about
5-8 million years ago. The shock-compressed gas then settled in two,
counter-rotating disks which probably subsequently lost further angular
momentum, moved inwards and finally became gravitationally unstable and
formed stars.

\item In contrast to the massive stars at $p$=1-10'', the early type stars
in the 'SgrA$^{\ast }$ cluster' (the 'S'-stars) in our opinion are most
likely formed as a result of collisions and mergers of lower mass stars, in
analogy to the blue straggler phenomenon in globular clusters.

\item The star closest to the black hole in 2002, S2, exhibits a
mid-infrared excess. We propose that this excess is caused either by
infrared emission from the accretion region around SgrA$^{\ast }$ itself, or
by the interaction of the star's UV light with dust in the accretion flow
onto the black hole. In the latter case, the lower limit to the accretion
rate 10-40 mas (1400-5700 Schwarzschild radii) from the hole is about 10$%
^{-7}$ M$_{\odot }$ yr$^{-1}$.
\end{enumerate}

\bigskip \acknowledgements

We are grateful to all members of the NAOS/CONICA team (from MPIA, MPE,
Observatoire de Grenoble, Observatoire de Meudon, ONERA and ESO-Garching)
and the ESO-Paranal staff whose hard and dedicated work made these
observations possible. In particular, we thank N.Ageorges, K.Bickert,
W.Brandner, E.Gendron, M.Hartung, N.Hubin, C.Lidman, A.-M. Lagrange, A.F.M.
Moorwood, C.R\"{o}hrle, G.Rousset and J.Spyromilio. We thank C.Bailyn,
M.Davies and A.Sills for discussions on the properties of blue stragglers,
C.McKee for comments on dust destruction in fast shocks, and Y.Levin,
A.Beloborodov and S.Nayakshin for interesting discussions about the stars in
the black hole environment. We thank M.Goss for making available to us his
high resolution VLA 1.3cm data. Our work is in part based on observations
obtained at the Gemini Observatory, which is operated by the Association of
Universities for Research in Astronomy, Inc., under a cooperative agreement
with the NSF on behalf of the Gemini partnership: the National Science
Foundation (United States), the Particle Physics and Astronomy Research
Council (United Kingdom), the National Research Council (Canada), CONICYT
(Chile), the Australian Research Council (Australia), CNPq (Brazil) and
CONICET (Argentina). Collaborative research between MPE and Tel Aviv
University is supported by the German-Israeli Foundation (grant
I-0551-186.07/97). TA is supported by GIF grant 2044/01, Minerva grant 8484
and a New Faculty grant by Sir H.Djangoly, CBE, London, UK.

\bigskip

\bigskip

\bigskip

\appendix

\section{Build-up of massive stars by collisions and mergers}

Following Alexander (1999), the time scale for the collision of a star of
mass $m_{1}$ with another star of mass $m_{2}$ ($M_{\odot }$), with stellar
radii $r_{1}$ and $r_{2}$, at an average radius R(arcsec) from SgrA$^{\star
} $ can be expressed as 
\begin{equation}
t_{c}=\left\{ 4\sqrt{\pi }\sigma C_{g}n_{2}(xr_{1}+r_{2})^{2}\left[ 1+\frac{%
C_{f}G(m_{1}+m_{2})}{C_{g}2(xr_{1}+r_{2})\sigma ^{2}}\right] \right\}
^{-1}\,,
\end{equation}%
where the normal geometric cross section (g) and gravitational focusing (f)
terms have been modified by average correction factors $C_{g}$ and $C_{f}$
to account for orbit averaging. The factor $x$ describes how off-center the
collision is relative to the center of star 1 ($x=0$ is central, $x=1$ is
grazing). We assume that the stellar density distribution is described by 
\begin{equation}
n_{\star }(R)=\frac{\rho _{\star }(R)}{\langle m\rangle }=2.2\times
10^{7}R_{1}^{-1.4}\,\,\,[M_{\odot }\mathrm{pc}^{-3}]\,,
\end{equation}%
where $\langle m\rangle =1.3$ $M_{\odot }$ is the average stellar mass for a
Salpeter mass function between 0.5 and 20 $M_{\odot }.$ R$_{1}$ is the
separation from SgrA$^{\ast }$ in arcseconds. To normalize the last
equation, we have taken a broken power law density distribution ($\rho
\varpropto R^{-\alpha }$) with an exponent $\alpha =1.4$ inside $10^{\prime
\prime }$ and 2 outside, and with the mass normalization discussed in
section 3.2 (Genzel et al. 1996). The density of collision partner $n_{2}$
then can be expressed as $n_{2}=n_{\star }f(m_{2})$, where $f(m)$ is the
fraction of stars in mass bin $m$. The one dimensional velocity dispersion
scales with radius from the black hole of mass M$_{\bullet }$=3x10$%
^{6}M_{\odot }$ as (Alexander 1999) 
\begin{equation}
\sigma (R)=v_{c}/\sqrt{1+\alpha }=\sqrt{\frac{GM_{\bullet }}{(1+\alpha )R}}%
=364\text{ }R_{1}^{-1/2}\,\,\,[\mathrm{km\,s}^{-1}]\,.
\end{equation}%
For main sequence stars of mass $m(M_{\odot })$ we take the standard
radius-mass relationship (Cox 2000) 
\begin{equation}
r(m)=1.03m^{0.64}\,\,\,[R_{\odot }]\,.
\end{equation}%
For $\alpha =1.4$ Alexander (1999) finds $C_{g}\sim $ 2.08 and $C_{f}\sim
1.36$, which reflects the fact that $\sim 80\%$ of the cusp stars are bound
to the black hole and thus, that they spend significant time at $R<\langle
R\rangle $ where the collisions are more frequent. For collisions to be
effective, they have to occur within the lifetime of the star $m_{1}$. We
take the mass dependence of the main sequence lifetime to be (Genzel et al.
1994) 
\begin{equation}
t_{ms}(m)=\left\{ 
\begin{array}{ll}
10^{9.8}m^{-2.5}\,\,\,[yr] & \mbox{for $m_{10}\leq0.9$} \\ 
10^{8.7}m^{-1.5}\,\,\,[yr] & \mbox{for $m_{10}>0.9$}%
\end{array}%
\right.
\end{equation}%
Collisions are effective for forming a star of mass $m_{3}$ from the merger
of $m_{1}$ and $m_{2}$ if 
\begin{equation}
P(m_{3},m_{2},m_{1})=\frac{\tau _{ms}(m_{3})}{t_{c}(m_{1},m_{2})}\geq 1\,.
\end{equation}%
For grazing collisions of two main sequence stars the collision time is 
\begin{equation}
t_{c}(m,m)=4\times
10^{9}m^{-1.28}R_{1}^{1.9}(1+0.5m^{0.36}R_{1})^{-1}f(m)^{-1}\,\,\,[\mathrm{yr%
}]\,.
\end{equation}

For $m=1$ $M_{\odot }$ the radius $R_{c}$ within which $P\geq 1$ then is
about $0.6^{\prime \prime }$. If the merging efficiency is high for grazing
collisions (c.f. references in Genzel et al. 1994), $R_{c}$ is then also the
radius within which many 1 $M_{\odot }$ stars continuously merge to 2 $%
M_{\odot }$ stars. The `S'-sources require collisions and merging of more
massive stars, however. For $m=5$ ($m_{3}=10\,M_{\odot }$ merger product) $%
R_{c}$ is about $0.3$ $f(5)^{0.53}$, or $\geq 0.1^{\prime \prime }$ for $%
f(5)\geq 0.1$. Obviously a more realistic calculation of the entire merger
tree is desirable but our estimate already suggests that the merger model
predicts a significant number of $\sim $10 $M_{\odot }$ stars to occur
within the SgrA$^{\star }$ cluster zone. Our estimates obviously require
that the merging efficiency is high and that the angular momentum of the
-initially- rapidly rotating merger remnant is removed quickly. Very little
is known theoretically about the merging efficiencies in high velocity
collision domain that characterizes the Galactic center but naively one
would expect that they should be low. However recent SPH simulations of
stellar collisions at high velocities by one of us (T.A.) suggest that
merging efficiencies may be surprisingly large under a fairly wide range of
conditions. How angular momentum is transported away in stellar mergers is
not understood quantitatively (Sills \& Bailyn 1999, Sills et al. 2001).
Qualitatively it is thought that some sort of magnetic breaking, either
through a wind/outflow, or through a locked circumstellar disk is efficient
(Sills et al. 2001). Blue stragglers are obviously formed in significant
numbers in globular clusters, indicating that nature has found a way to
solve the angular momentum problem (Bailyn 1995). In the one case where a
rotation velocity is measured (Shara, Saffer \& Livio 1997), the rotation is
moderately fast (v$_{rot}$sin (i)=155 km/s) but significantly below breakup.
The deduced rotation velocity of S2 (224 km/s) is also not unusual for a
normal late O star (Ghez et al. 2003).

Another predicted effect of the collision model in the cusp (Alexander \&
Livio 2001; Alexander \& Kumar 2001; Alexander 2002; Alexander \& Morris
2003) is the destruction of late type giants . Again we apply the above
equations with $m_{1}=m_{g}\sim 2$ $M_{\odot }$, $m_{2}=\langle m\rangle ,$
and assume that near-central collisions ($x\sim 0.25$) between a giant and a
low mass, dwarf star result in the permanent destruction of the giant's
envelope (Davies et al. 1998, Alexander 1999). Bright red giants (M1 to K5: K%
$_{s}\sim 12-14$) have a lifetime of about 10\% of $\tau _{ms}$ and a radius
of 20-80 $R_{\odot }$ (Cox 2000). The collision time then is 
\begin{equation}
t_{c}(g,\langle m\rangle )\sim 9\times
10^{7}R_{1}^{2}(1+0.1R_{1})^{-1}f(\langle m\rangle )^{-1}\,\ [\text{\textrm{%
yr}}],
\end{equation}%
and $P\geq $1 occurs within $R_{c}\sim 1.2^{\prime \prime }$ for a 50 $%
R_{\odot }$ giant. In this regime $R_{c}$ scales approximately with $%
(xr_{g})^{2}$ so that destruction occurs at even greater radii for the even
larger AGB stars and if the envelope is destroyed at more off-center
collisions. We have shown above (Fig. 7) that the fraction of K$_{s}\leq 15$
giants decreases within the central $2^{\prime \prime }$, in good agreement
with the collider model and the earlier conclusions of Alexander (1999).

\bigskip

\bigskip

\section{Heating of dust in the accretion flow by UV radiation from S2}

We show here that the mid-IR excess emission of S2/SgrA$^{\ast }$ may come
from dust in the accretion flow (onto the black hole) that is heated by the
hot and luminous O star S2. The UV dust optical depth of this flow can be
estimated from 
\begin{equation}
\tau _{\mathrm{UV}}=f_{\mathrm{dust}}\left( \frac{\dot{M}r/4\pi R^{2}v\mu }{%
6.3\times 10^{20}\mathrm{cm}^{-2}}\right) \,.
\end{equation}%
Here $\dot{M}$ is the mass accretion rate onto SgrA$^{\star }$, R ($\sim
2\times 10^{-3}$ pc) is the distance of S2 from SgrA$^{\star },$ $\mu $ is
the mean molecular weight (3.4x10$^{-24}$ g), $v$ is the velocity of the
accretion flow, $r$ is the radius of the dust column from S2, and $f_{%
\mathrm{dust}}$ is the fraction of dust (relative to the normal interstellar
medium) that is not destroyed due to the accretion shock in the hot
accretion flow near S2. The temperature the dust particles attain is given
from the heating-cooling balance, 
\begin{equation}
T_{d}(r)\sim T_{o}\left( \frac{L_{UV}}{L_{o}}\right) ^{1/(4+\beta )}\left( 
\frac{r}{r_{o}}\right) ^{-2/(4+\beta )}\,.
\end{equation}%
For dust particles with wavelength dependent extinction efficiency $%
Q(\lambda )=Q(\lambda _{o})(\frac{\lambda }{\lambda _{o}})^{-\beta }$ and
normal interstellar dust parameters, $\beta =1$, $T_{o}=82K$, $%
L_{o}=10^{5}L_{\odot }$ and $r_{o}=3\times 10^{17}$ cm (Scoville \& Kwan
1976). For simplicity we assume further that the emission at $\lambda $ is
dominated by those dust particles whose temperature at $r$ is equal to the
Wien temperature $(T_{\mathrm{Wien}}(L^{\prime })=710\,\mathrm{K})$. For a
late O star the UV luminosity is about half of the bolometric luminosity,
such that $r(T_{\mathrm{Wien}}(L^{\prime }))=10^{15}$ cm. The emitted flux
density of this hot dust cloud around the O/B star then is 
\begin{equation}
S(L^{\prime })=17.5\text{ }r_{15}^{2}\tau _{UV}\,\ \ \ [Jy]\,,
\end{equation}%
for an assumed 8 kpc distance to the Galactic Center. Comparison to the
observed flux density implies $\tau _{UV}\sim 1.7\times 10^{-3}$. For $%
v=10^{3\text{ }}$km/s we find 
\begin{equation}
\left[ \dot{M}f_{\mathrm{dust}}\right] _{R=0.002\mathrm{pc}}\sim 2\times
10^{-7}\,\ [M_{\odot }\,\mathrm{yr}^{-1}]\,.
\end{equation}

\bigskip Dust destruction in fast shocks is mostly due to sputtering. It
strongly depends on the pre-shock gas density and leads to a deficiency of
small grains (Dwek, Foster \& Vancura 1996). For the gas densities in the
pre-accretion shock region in the Galactic Center (between 0.1 to 10 cm$%
^{-3} $ at R=1'') we estimate from the models of Dwek et al. that 15 to 70\%
of the dust is destroyed.



\bigskip

\clearpage

\begin{table}[tbp]
\begin{center}
\begin{tabular}{c|ccc}
$K_{inp}$ & $K_{recov}$ & $\sigma$ & Completeness [\%] \\ 
\tableline 13 & 13.05 & 0.03 & 99 \\ 
14 & 14.04 & 0.05 & 98 \\ 
15 & 15.04 & 0.07 & 95 \\ 
16 & 16.04 & 0.09 & 88 \\ 
17 & 17.04 & 0.11 & 79 \\ 
18 & 18.01 & 0.17 & 63 \\ 
19 & 18.91 & 0.22 & 32 \\ 
20 & 19.76 & 0.23 & 4 \\ 
\tableline &  &  & 
\end{tabular}%
\end{center}
\caption{Average magnitude of recovered sources, sigmas of average
magnitudes and overall completeness for given magnitudes.}
\end{table}

\clearpage

\begin{table}[tbp]
\begin{center}
{\small 
\begin{tabular}{c|ccc|ccc|}
magnitude limit & $\beta$ =0.21: $p \leq 1"$ & $\leq 0.1"$ & $\leq 0.03"$ & $%
\beta=0.35: p \leq 1"$ & $\leq 0.1"$ & $\leq 0.03"$ \\ 
\tableline K$_{s}\leq$ 18.5: A0V/3.5 $M_{\odot }$ & 114 & 3 & 0.4 & 189 & 4
& 0.6 \\ 
K$_{s}\leq$ 19.5: A5V/2.2 $M_{\odot }$ & 185 & 4 & 0.7 & 423 & 10 & 1.4 \\ 
K$_{s}\leq$ 21: G1/2V/1.0 $M_{\odot }$ & 383 & 9 & 1.3 & 1416 & 33 & 5 \\ 
K$_{s}\leq$ 23: M0V/0.5 $M_{\odot }$ & 1000 & 24 & 3.4 & 7100 & 166 & 23%
\end{tabular}
}
\end{center}
\caption{Numbers of stars expected from our mass/KLF modelling, for
different magnitude/mass limits, for two values of the KLF faint-end slope $%
\protect\beta$, and for three different outer radii from SgrA$^\star$.}
\end{table}

\clearpage

\begin{table}[tbp]
\begin{center}
\begin{tabular}{c|ccccccccccc}
source & $x(\mathrm{SgrA}^\star)$ & $y(\mathrm{SgrA}^\star)$ & $H$ & $\delta
H$ & $K_{s}$ & $\delta K_{s}$ & $L$ & $\delta L^{\prime}$ & H-K$_{s}$ & K$%
_{s}$-L' &  \\ \hline
SgrA$^\star$ & 0 & 0 & $>$18.4 & 0.3 & $>$16.6 & 0.3 & 11.7 & 0.5 &  &  & 
\\ 
S2+SgrA$^\star$ & 0.02 & 0.04 & 15.36 & 0.1 & 14.01 & 0.1 & 11.12 & 0.1 & 1.3
& 2.9 &  \\ 
S1 & -0.05 & -0.18 & 16.1 & 0.1 & 14.79 & 0.12 & 12.86 & 0.2 & 1.3 & 1.9 & 
\\ 
S4 & 0.25 & 0.13 & 15.74 & 0.1 & 14.42 & 0.1 & 12.60 & 0.13 & 1.3 & 1.8 & 
\\ 
S12 & -0.08 & 0.27 & 16.88 & 0.1 & 15.58 & 0.1 & 13.99 & 0.15 & 1.3 & 1.6 & 
\\ 
S9 & 0.17 & -0.34 & 16.55 & 0.1 & 15.17 & 0.12 & 12.95 & 0.1 & 1.4 & 2.2 & 
\\ 
S10 & 0.050 & -0.37 & 15.78 & 0.1 & 14.23 & 0.1 & 12.14 & 0.1 & 1.5 & 2.1 & 
\\ 
S8 & 0.36 & -0.24 & 15.89 & 0.1 & 14.52 & 0.1 & 12.63 & 0.12 & 1.4 & 1.9 & 
\\ 
S5 & 0.46 & 0.10 & 16.92 & 0.1 & 15.44 & 0.12 & 13.29 & 0.2 & 1.5 & 2.2 & 
\\ 
S7 & 0.51 & -0.03 & 16.70 & 0.11 & 15.34 & 0.14 & 12.63 & 0.13 & 1.4 & 2.7 & 
\\ 
S11 & 0.15 & -0.55 & 15.78 & 0.1 & 14.43 & 0.1 & 12.43 & 0.12 & 1.4 & 2.0 & 
\\ 
IRS16SWNW & 0.53 & -0.43 & 14.88 & 0.1 & 13.41 & 0.1 & 11.65 & 0.1 & 1.5 & 
1.8 &  \\ 
& 0.51 & 0.45 & 16.37 & 0.1 & 15.13 & 0.1 & 13.20 & 0.11 & 1.2 & 1.9 & 
\end{tabular}%
\end{center}
\caption{HK$_{s}$L' magnitudes in Aug 2002 of H$<\!17$ sources and SgrA$%
^\star$ in the central $0.7^{\prime\prime}$.}
\end{table}

\clearpage 

\begin{figure}[tbp]
\plotone{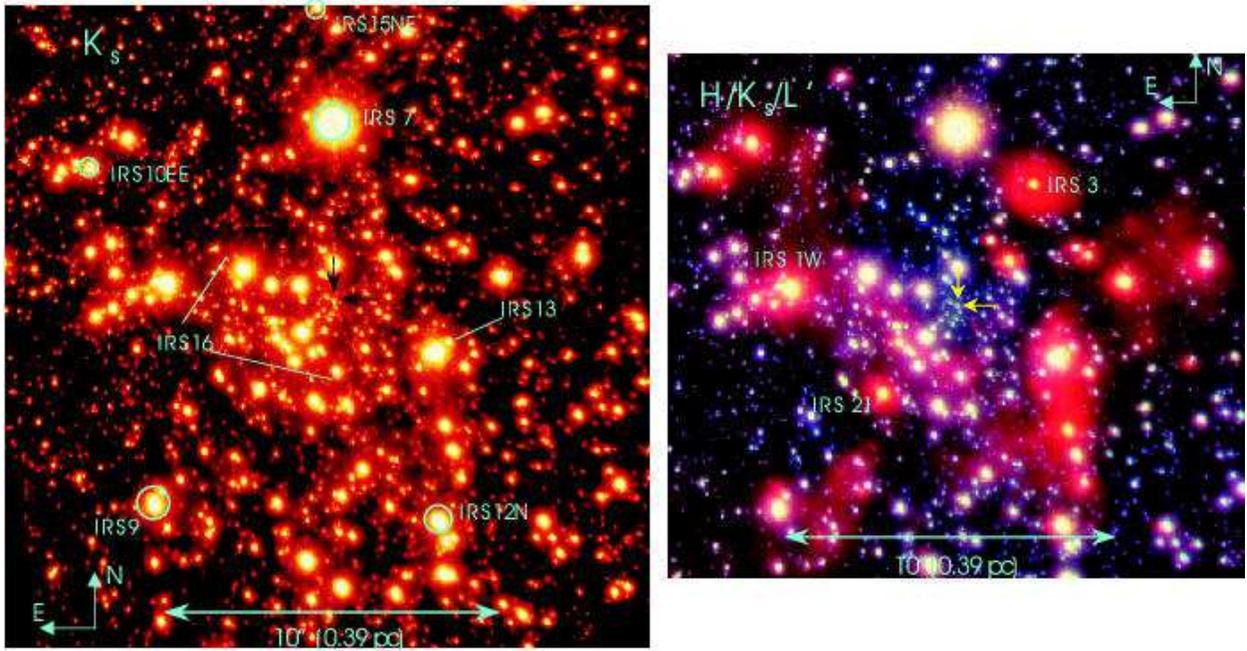}
\caption{ Left: K$_{s}$-band NAOS-CONICA shift-and-add image of the central $%
\sim 20^{\prime \prime }$, taken in August 2002 (55 mas FWHM resolution,
Strehl ratio $\sim 50\%$). A logarithmic color scale is used. East is to the
left, and north is up. The brightest star, IRS7 (6.7 mag) was used for
infrared wavefront sensing. The faintest stars visible on the image are K $%
_{s}\sim $ 19. The five encircled stars (of 7 in the central 30$^{\prime
\prime }$) are also radio SiO masers and were used for establishing the
selected orientations of the infrared camera and to put the infrared data in
the radio astrometric reference frame ( rms$\sim \pm $10 mas, Reid et al.
2002). In addition the IRS16 and IRS13 complexes of emission line stars are
marked. The two arrows denote the position of SgrA$^{\star }$. Right: H/K$%
_{s}$/L$^{\prime }$ three color composite of the central $\sim 20^{\prime
\prime }$. East is to the left, and north is up. Several bright, dusty
L-band excess stars are marked, and the two arrows denote the position of
SgrA$^{\star }$.}
\end{figure}

\clearpage

\begin{figure}[tbp]
\epsscale{0.8} \plotone{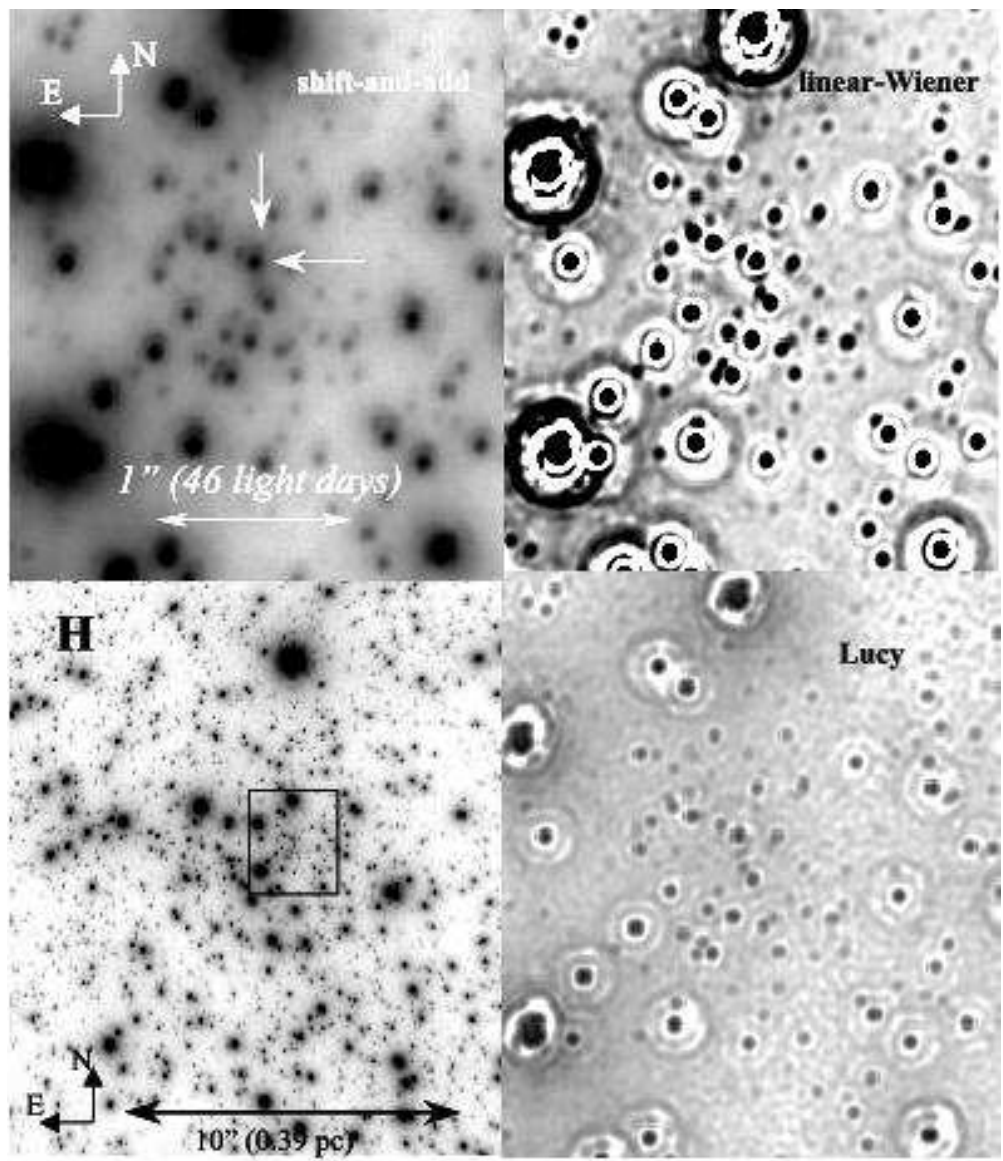}
\caption{ Bottom left: grey scale H-band shift-and-add (SSA) image of the
central $\sim 20^{\prime \prime }$ region. A logarithmic color scale is
used. The spatial scale is marked, North is up and East is to the left.
Resolution is 40 mas (FWHM) and the Strehl ratio is about 33\%. Top left:
Zoom of SSA image into the central 2-3$^{\prime \prime }$ region surrounding
SgrA$^{\star }$ (marked as a rectangle in the lower left image). The
position of SgrA$^{\star }$ is marked by arrows. Bottom and top right: Lucy
deconvolved (bottom) and linearly deconvolved/Wiener filtered (top),
processed versions of the SSA image in the upper left (same spatial scale
and logarithmic color scale). In the case of the Lucy image the final `$%
\protect\delta $'-function map was reconvolved with a Gaussian of 40 mas
FWHM. The main purpose of the deconvolution is to remove the seeing halos,
and not to enhance the spatial resolution of the images. Rings in both
images are artefacts of the image processing. With the exception of the
vicinity of very bright, and partially saturated stars SSA, linearly and
Lucy deconvolved images show the same distribution of faint stars. }
\end{figure}

\clearpage

\begin{figure}[tbp]
\epsscale{1.0} \plotone{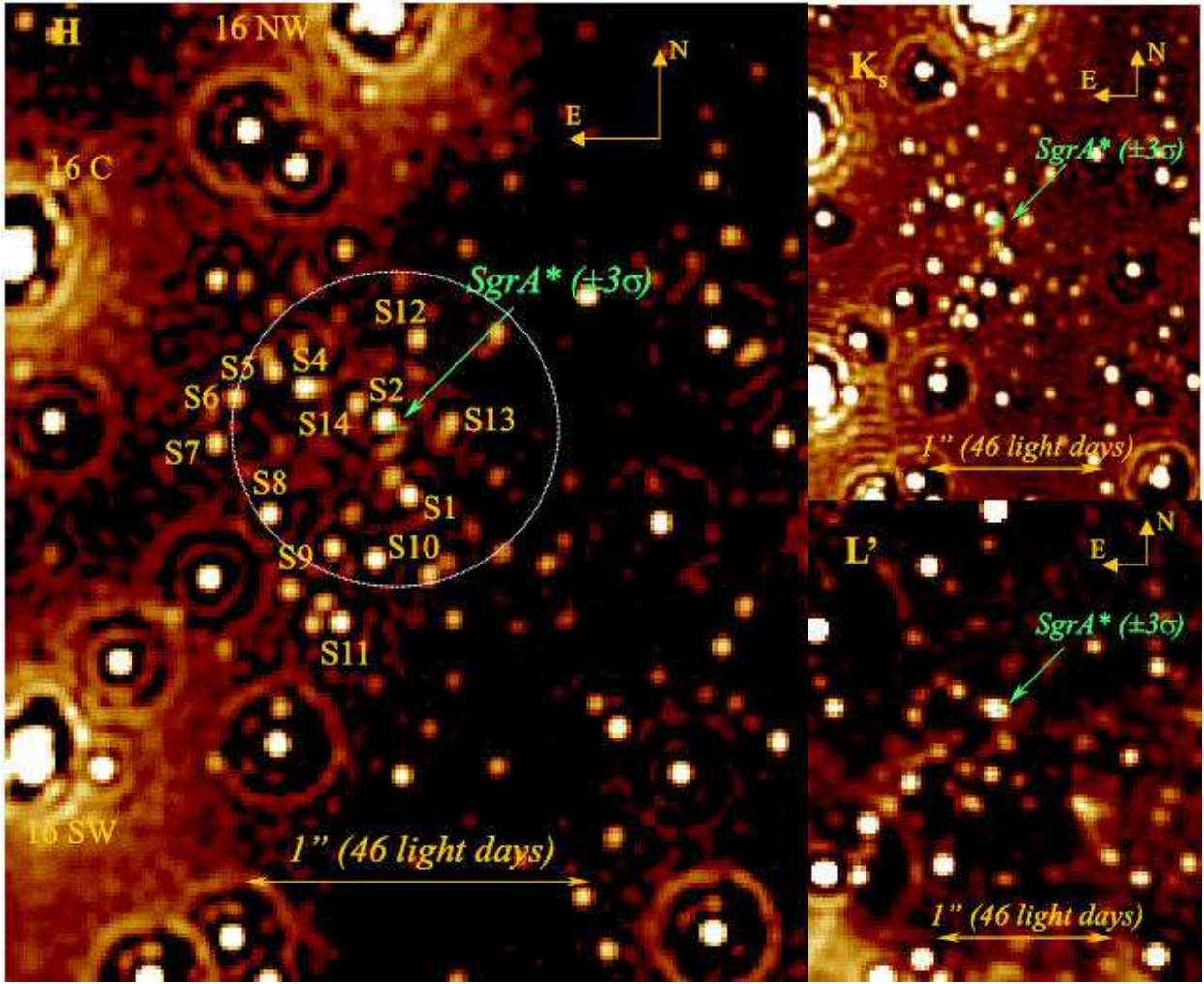}
\caption{ Lucy-Richardson deconvolved images at H (left, reconvolved to 40
mas FWHM resolution), K$_{s}$(top right, reconvolved to 55 mas resolution)
and L$^{\prime }$ ( bottom right, reconvolved to 70 mas resolution) of the
same central 2$^{\prime \prime }$ region around SgrA$^{\star }$. The
position of SgrA$^{\star }$ (and its 3$\protect\sigma $ positional
uncertainty) is marked by a small cross. Several of the `S'-sources in the
SgrA$^{\star }$-cluster, and three bright IRS16 complex stars are marked on
the H-band image. The thin dashed circle marks the radius 0.5'', SgrA$^{\ast
}$ cluster region, containing the fastest moving stars tightly bound to the
black hole (S1, S2, S12, S13, S14). Note that for the L$^{\prime }$-band
image the visible wavefront sensor was used, tracking on a star $\sim $20$%
^{\prime \prime }$ northeast of SgrA$^{\star }$. The rings around brighter
stars are artifacts of the deconvolution. The faintest stars recognizable on
the images are H$\sim $20, K$_{s}\sim $18.5 and L$^{\prime }\sim 14$. }
\end{figure}

\clearpage

\begin{figure}[tbp]
\epsscale{0.5} \plotone{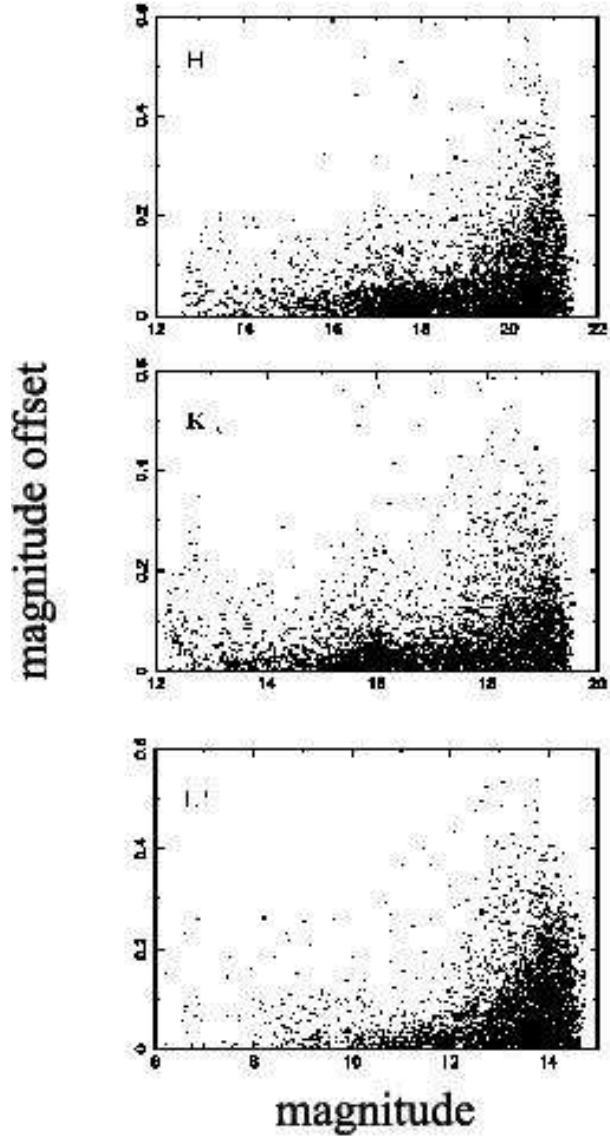}
\caption{ Photometric errors (in mag) as a function of magnitude as
determined from a comparison of the photometry obtained on the
Lucy-Richardson and Wiener deconvolved images in H (top), K$_{s}$ (middle)
and L$^\prime$ (bottom). }
\end{figure}

\clearpage

\begin{figure}[tbp]
\epsscale{1.0} \plotone{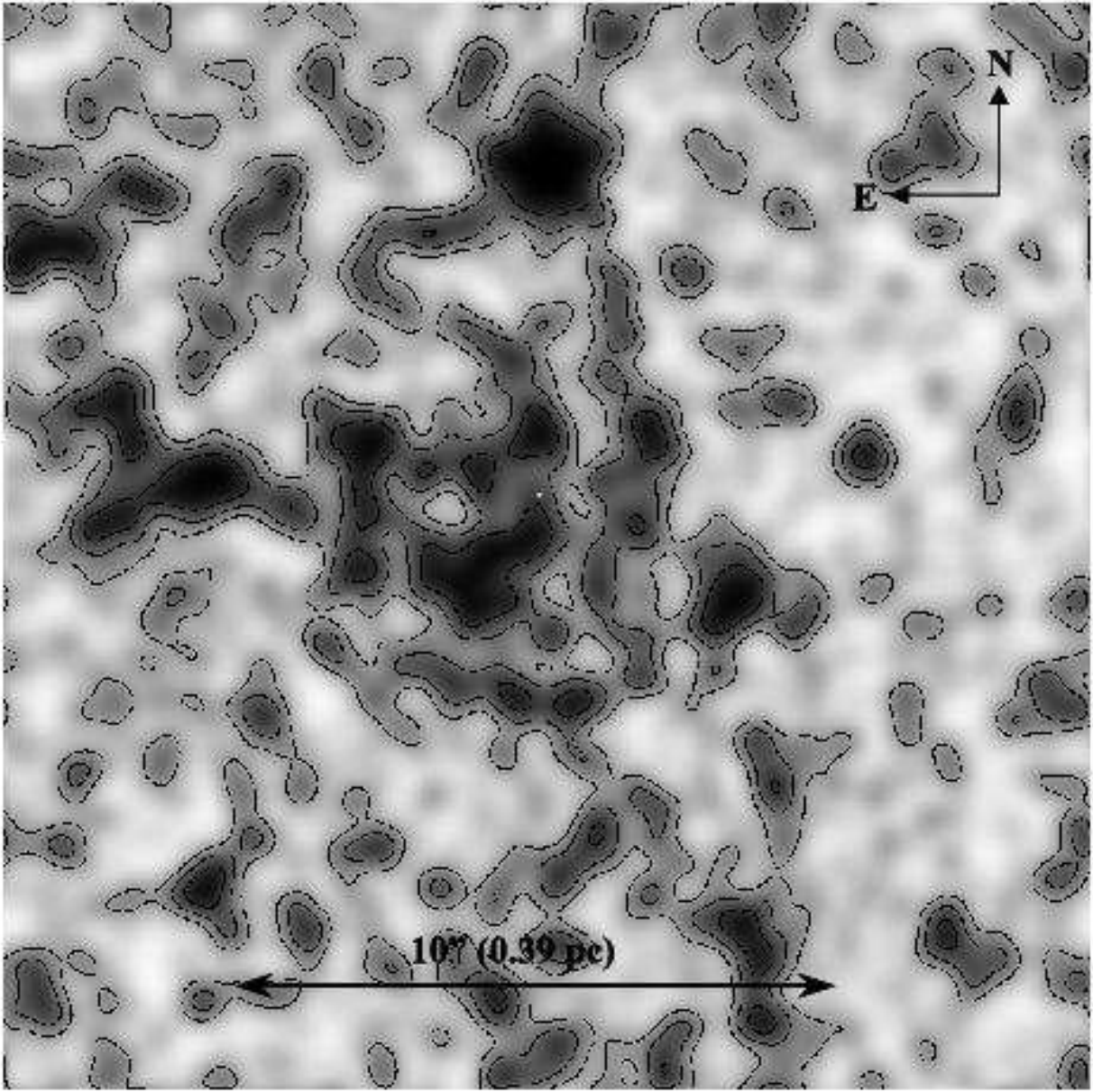}
\caption{ Completeness map of the K$_{s}$-image at K$_{s}=$ 18, obtained by
adding artificial K$_{s}=$ 18 stars to the image and determining the
probability of recovering a star within $\pm $0.5 mag of the input star. The
shades of grey correspond to the probability of recovering a source at a
given position. Dark areas indicate the lowest probabilities. The contours
outline the 20, 40 and 60\% probability levels. }
\end{figure}

\clearpage

\begin{figure}[tbp]
\plotone{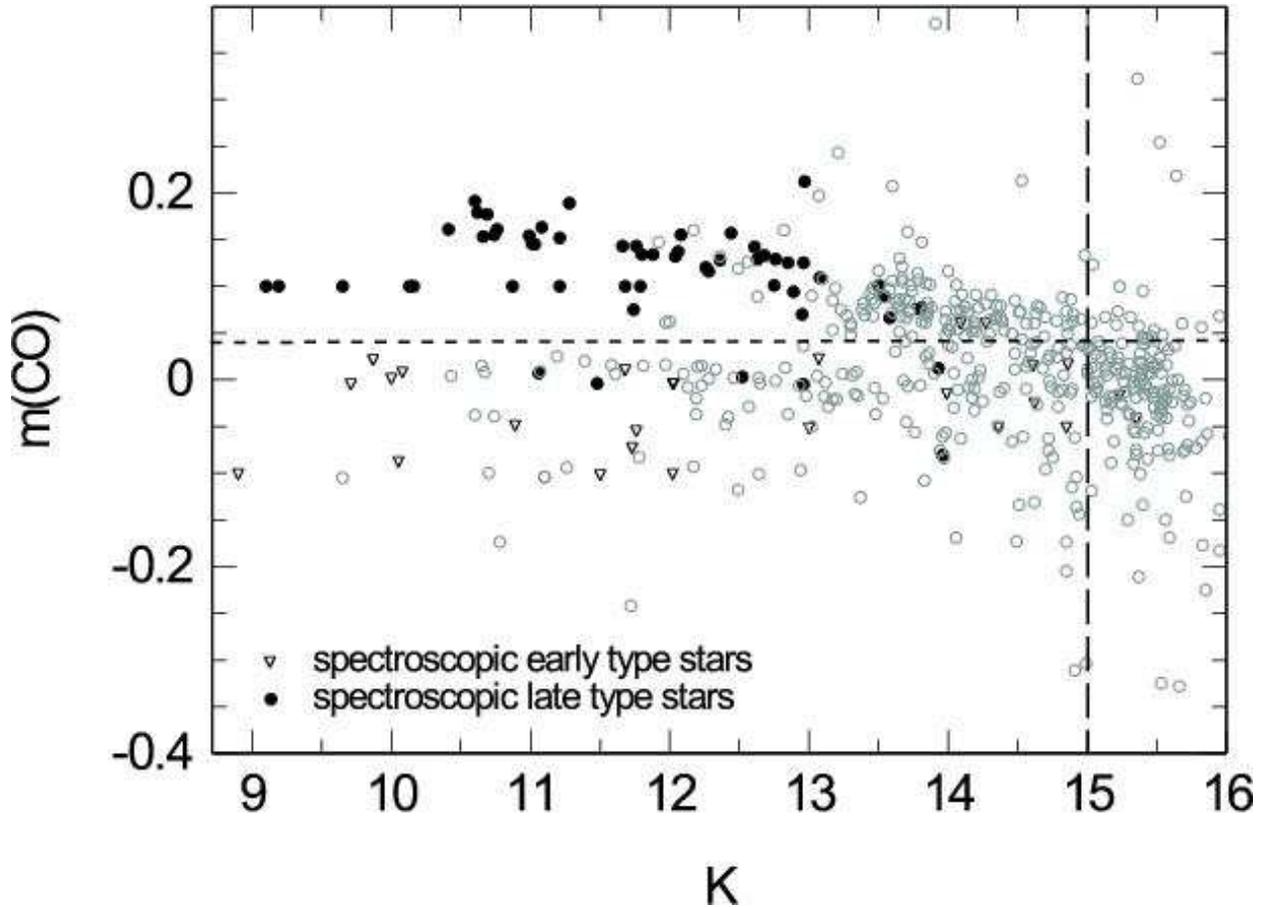}
\caption{ CO-index ($m(\mathrm{CO})=m(2.29)-m(2.26)$ ) as a function of
K-magnitude for those stars in the Ott et al. (2003) proper motion sample
that also have Gemini science demonstration data, narrow band maps (open
circles). Stars marked with triangles denote early type stars, and
stars marked with filled circles denote late type stars confirmed by the Ott
et al. (2003) spectroscopic data. For K$\leq 15$ stars with $m(\mathrm{CO}%
)\geq 0.04 $ are identified as late type stars, while stars with $m(\mathrm{%
CO}) < 0.04$ are identified as early type stars. }
\end{figure}

\clearpage

\begin{figure}[tbp]
\plotone{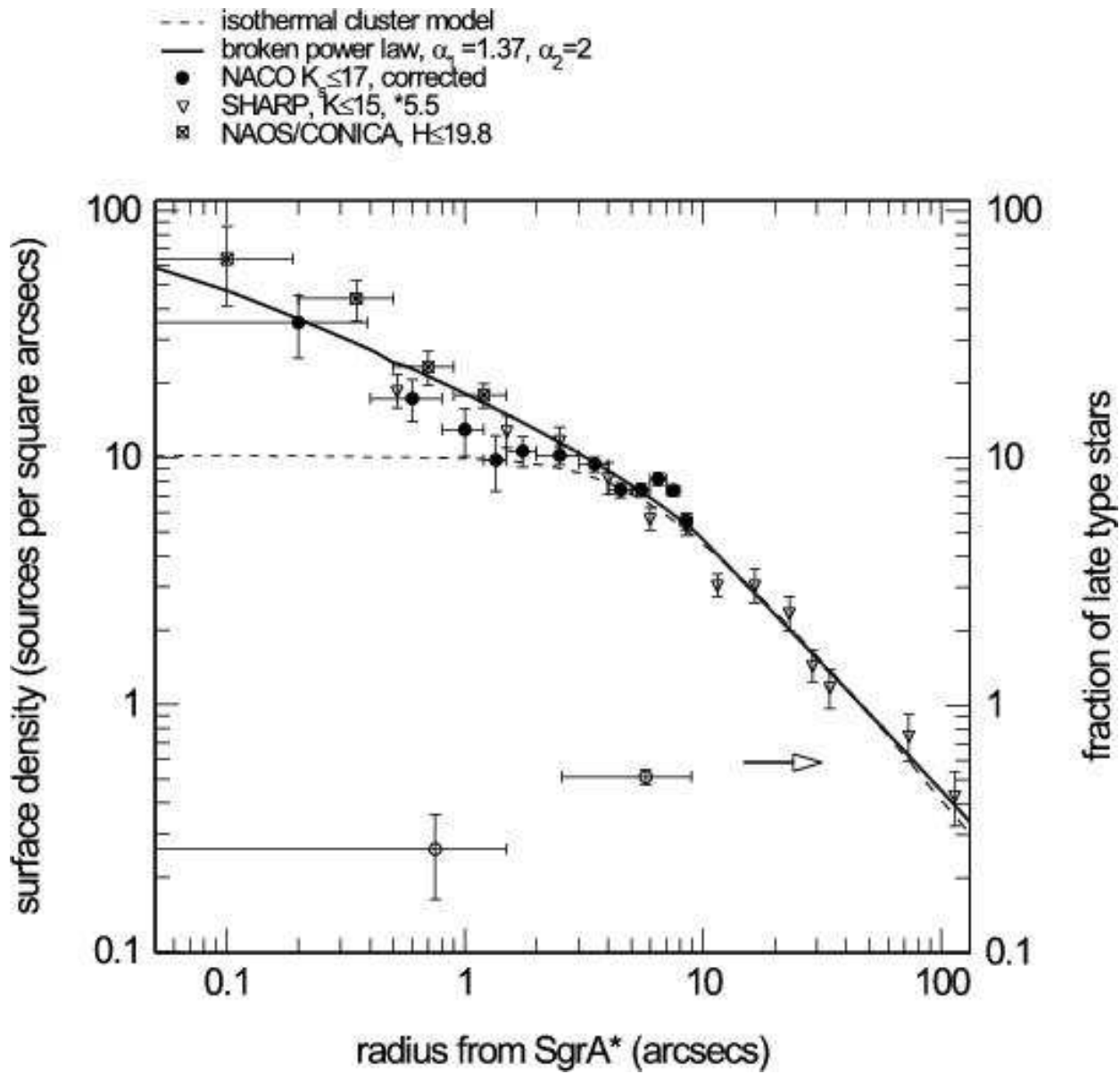}
\caption{ Surface density of stars as a function of projected radius from
SgrA$^{\star }$. Filled circles are NAOS/CONICA counts of all sources
present on both H and K$_{s}$ maps and with K$_{s}\leq $ 17, corrected for
incompleteness by the artificial star technique. Squares with crosses denote
direct H-band NAOS/CONICA counts by eye to H$\leq $19.8 near Sgr A*, a
region devoid of bright stars. Downward-pointing triangles denote the
SHARP K$\leq $ 15 counts from Genzel et al. (2000), and multiplied by a
factor of 5.5 to best match the deep NAOS/CONICA counts in the overlap
region beyond a few arcseconds from SgrA$^{\star }$. The dashed curve is the
model of a flattened isothermal sphere of core radius 0.34 pc fitting the
counts from SHARP data. Note that at K$\sim $15 the SHARP counts in the
innermost region are only 50\% complete. The continuous curve is the broken
power-law ($\protect\alpha $=2 beyond 10$^{\prime \prime }$ and $\protect%
\alpha $=1.4 within $10^{\prime \prime }$) discussed in the text. Open
circles at the bottom of the figure denote the fraction of late type stars
of the total K$\leq $ 15 sample with proper motions and Gemini CO
narrow-band indices. All vertical error bars are $\pm 1\protect\sigma $, and
denote the total uncertainty due to Poisson statistics and, where
appropriate, due to incompleteness/confusion correction ($\sim $10$\%$) in
each annulus. Horizontal bars denote the width of the annulus. }
\end{figure}

\clearpage

\begin{figure}[tbp]
\plotone{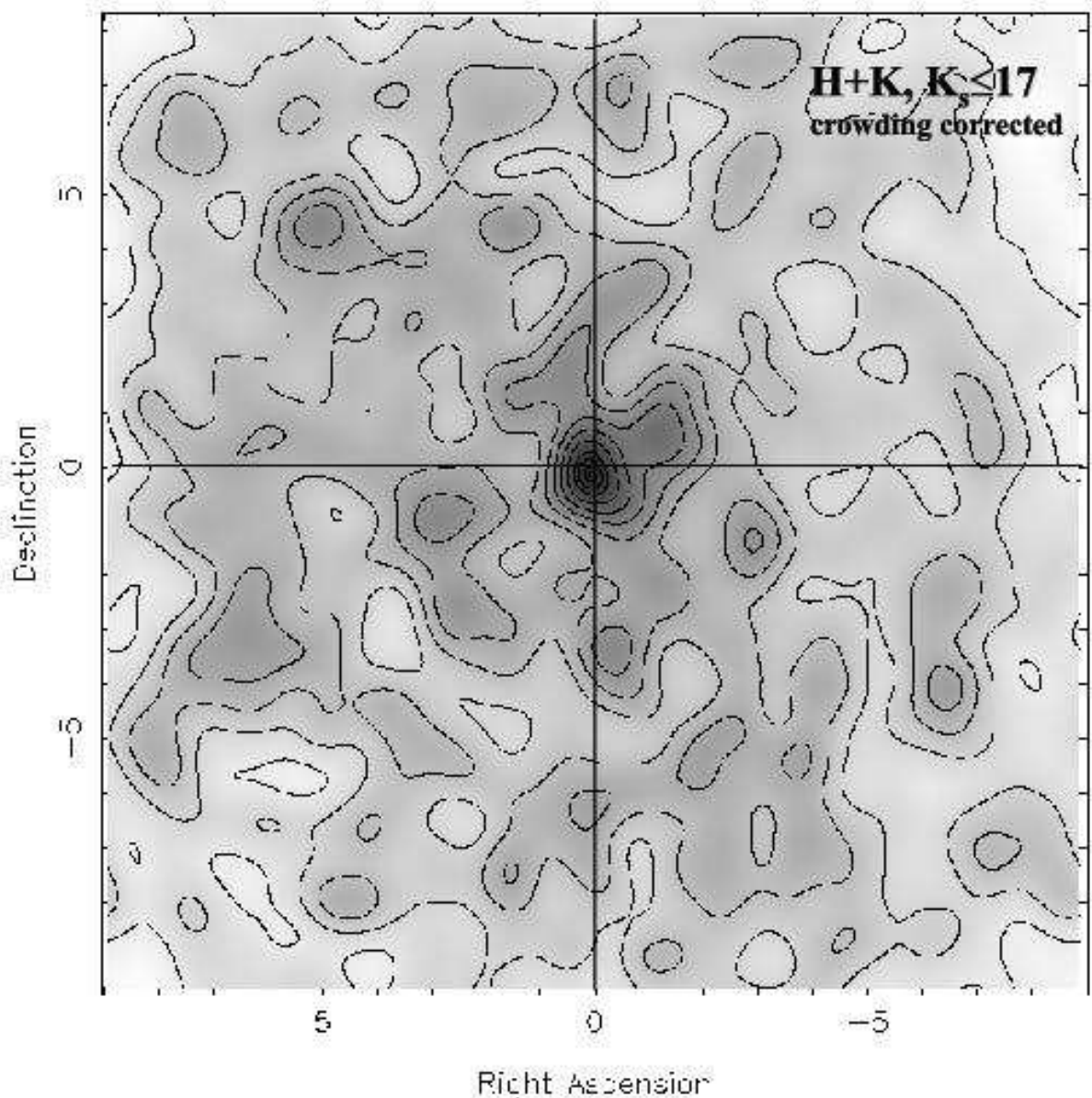}
\caption{ Two-dimensional map (in arcseconds relative to the map of Sgr A*)
of the smoothed surface density of NAOS/CONICA sources present on both H and
K$_{s}$ maps corrected for crowding and incompleteness. For this purpose,
the original source list maps of H+K sources in each magnitude bin between K$%
_{s} $ of 13 and 17 were smoothed with a 1$^{\prime\prime}$ Gaussian and
divided by the incompleteness map. The corrected maps of different
magnitudes were then added. Contours are 10, 20,\ldots,90, 95, 99 and 99.9\%
of the peak surface density. The maximum of the stellar density is at
(RA,Dec)=(0.09$^{\prime\prime}$, -0.15$^{\prime\prime}$) relative to SgrA$%
^\star$, with an uncertainty of $\pm $0.2$".$ }
\end{figure}

\clearpage

\begin{figure}[tbp]
\plotone{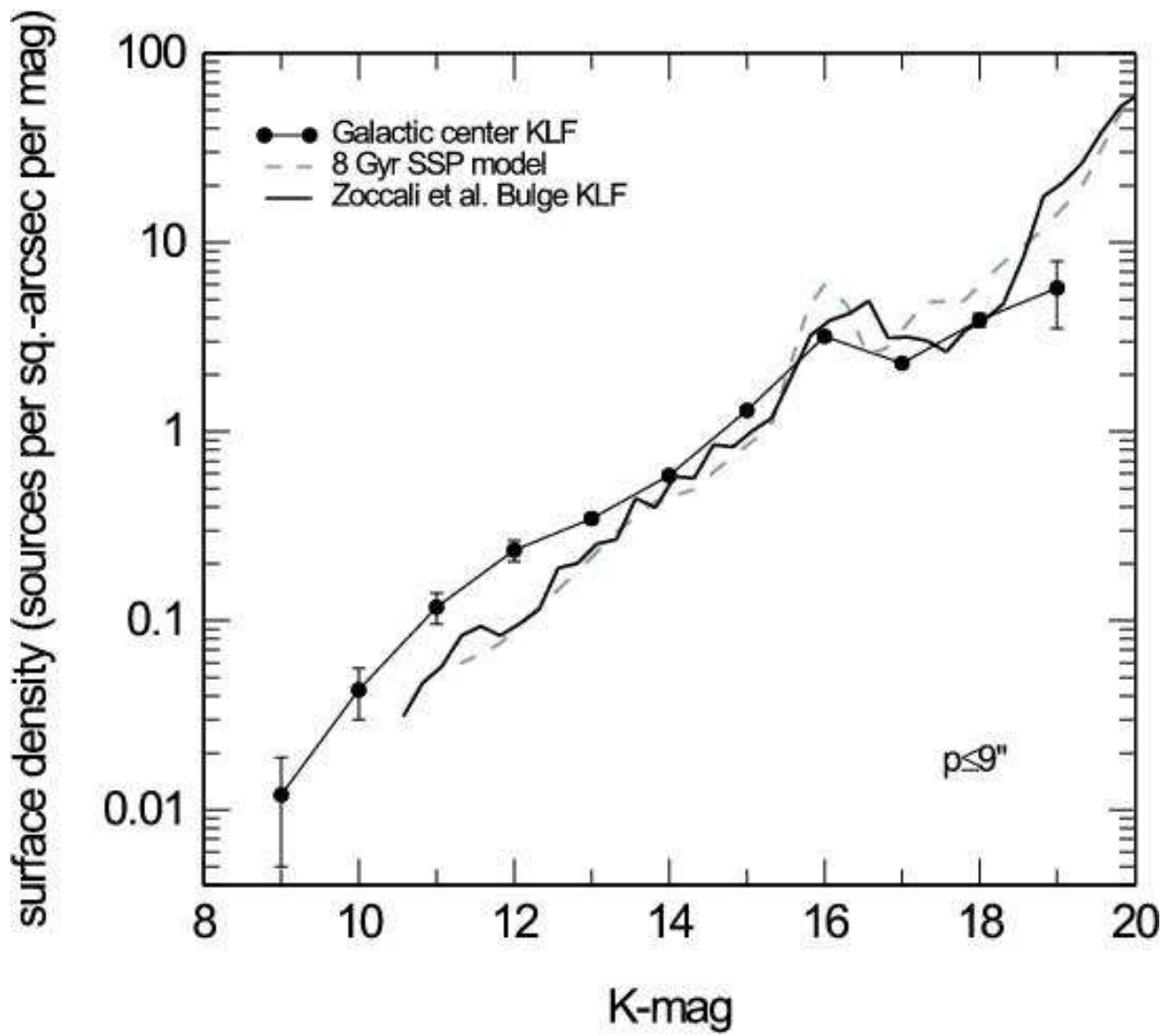}
\caption{ K-band luminosity function (sources per square arcsecond per mag)
as a function of K of the central 9$^{\prime \prime }$ region (filled
circles with $1\protect\sigma $ error bars). The KLF points have been
corrected for incompleteness as discussed in the text. Errors take into
account both the Poisson error, as well as the uncertainty of the
crowding/confusion correction ($\sim $10\%). The Galactic Center data points
are a combination of the new August 2002, NAOS/CONICA data (for K$_{s}\geq $
12) and of the SHARP/NTT data sets (Ott et al. 2003), scaled in the 12$\leq $
K$_{s}\leq $14.5 overlap region for best match with the CONICA data. For
comparison, the continuous curve is the KLF of the Galactic Bulge on scales
of degrees and the dashed curve is a single age (8Gyr) stellar population
model of the bulge (from Zoccali et al. 2002), both scaled vertically to
match the center data, and corrected horizontally to the same K-band
extinction. The prominent excess hump at K$_{s}\sim $ 16 is due to old metal
rich, low mass stars on the horizontal branch/red clump. }
\end{figure}

\clearpage

\begin{figure}[tbp]
\plotone{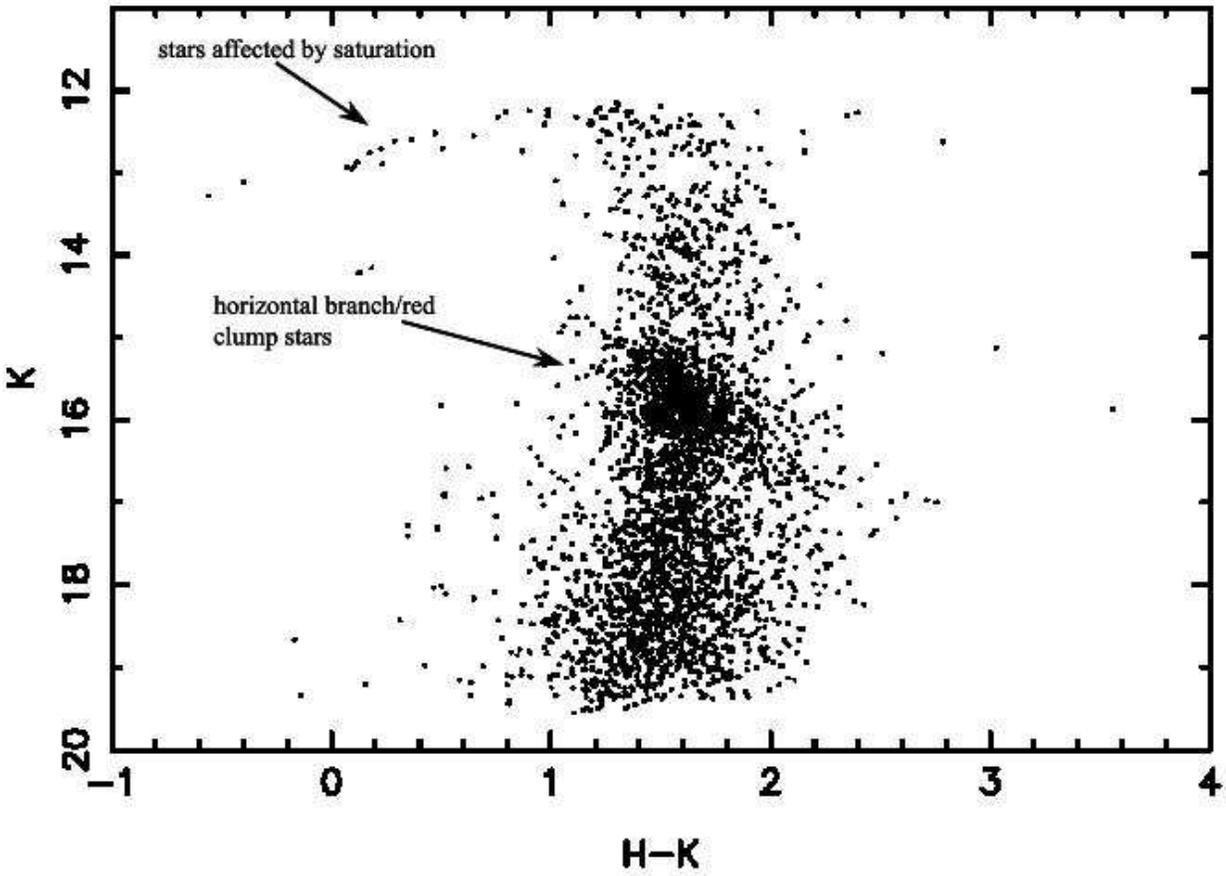}
\caption{ Color-magnitude diagram for the same $p\leq 9"$ region as in
Fig.9, again showing prominently the horizontal branch/red clump stars. Most
of the very blue stars in the color-magnitude at K$_{s}<13$ are probably
caused by saturation in the K$_{s}$-band image, which results in their K$%
_{s} $ magnitudes being underestimated. }
\end{figure}

\clearpage

\begin{figure}[tbp]
\plotone{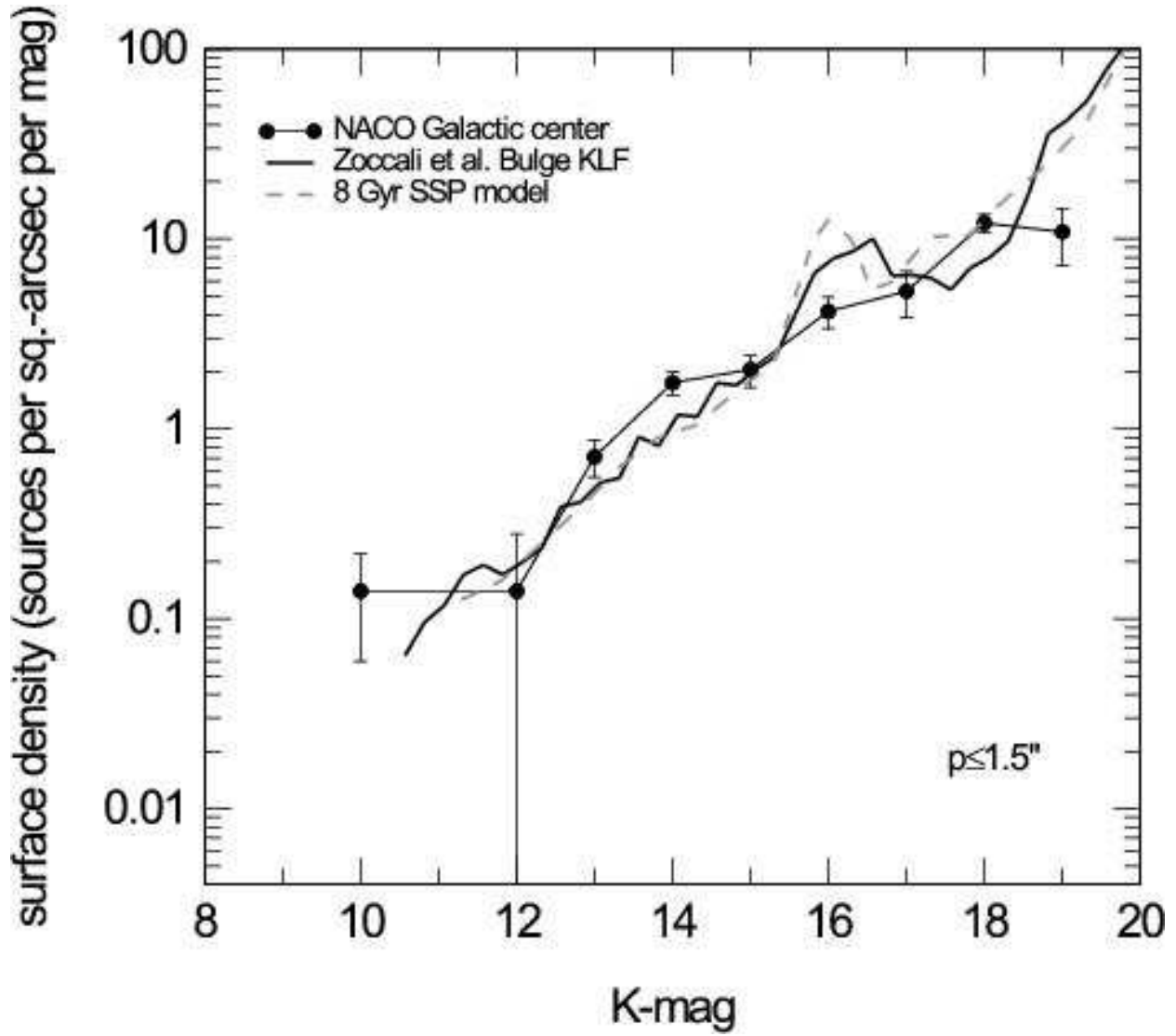}
\caption{ K-band luminosity function (sources per square arcsecond per mag)
as a function of K$_{s}$ of the central cusp region (filled circles, $\leq $%
1.5$^{\prime \prime }$,$\pm 1\protect\sigma $ error bars). Other symbols and
curves are as in Fig. 9.}
\end{figure}

\clearpage

\begin{figure}[tbp]
\plotone{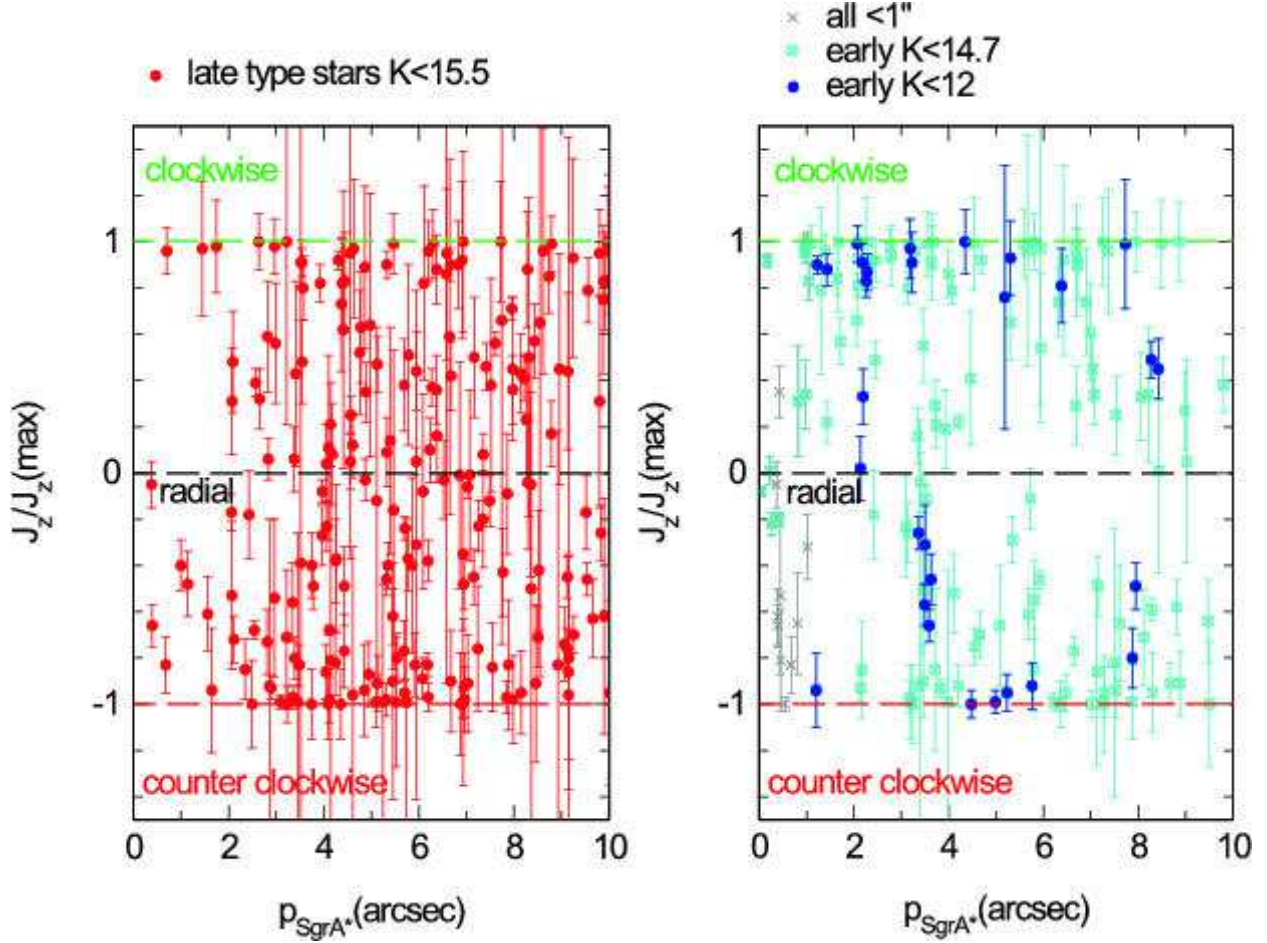}
\caption{ Normalized angular momentum along the line of sight ($%
J_{z}/J_{z}(\max )$) as a function of projected separation from SgrA$^{\star
}$ for the K$\leq 15.5$ late type stars in the left inset(filled circles: $m(%
\mathrm{CO})\geq 0.04$, left panel), and in the right inset for the early
type stars ($m(\mathrm{CO})<0.04$): K$\leq $14.7 (squares with crosses) and K%
$\leq $12 (filled circles), all from the Ott et al. (2003) sample. In
addition crosses denote the other $p\leq 1^{\prime \prime }$ stars in the
Ott et al. sample whose spectral properties are not known. In this plot,
clockwise projected tangential orbits are at $J_{z}/J_{z}(\max )=+1$,
counter-clockwise projected tangential orbits are at $J_{z}/J_{z}(\max )=-1$%
, and radial orbits are at $J_{z}/J_{z}(\max )=0$. All error bars are $\pm 1%
\protect\sigma $. }
\end{figure}

\clearpage

\begin{figure}[tbp]
\plotone{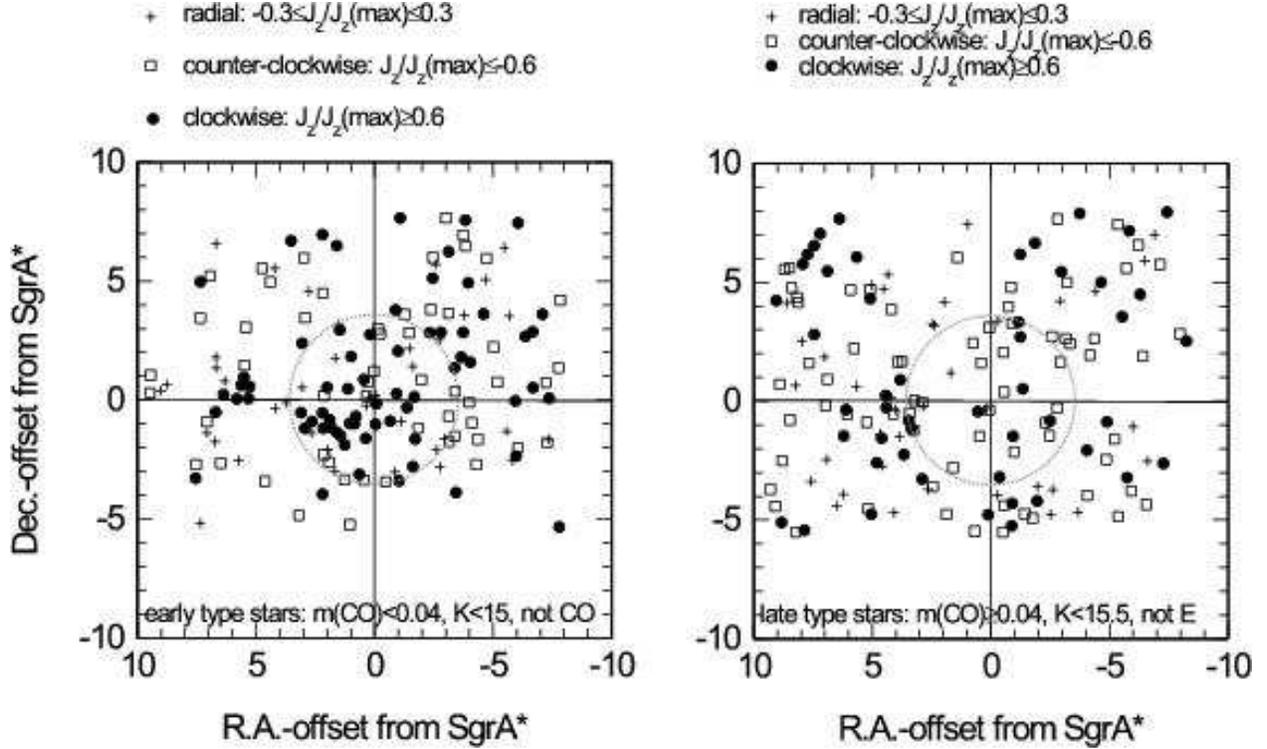}
\caption{ Spatial distribution of different projected orbits for early type
stars (left panel) and late type stars (right panel) in the Ott et al.
(2003) sample (for $K\leq 15$, and cutting at $m(\mathrm{CO})=0.04$, as in
Fig. 8). Filled circles denote projected clockwise tangential orbits ($%
J_{z}/J_{z}(\max )\geq 0.6$), open squares denote projected
counter-clockwise tangential orbits ($J_{z}/J_{z}(\max )\leq -0.6$), and
plus symbols denote projected radial orbits ($-0.3\leq J_{z}/J_{z}(\max
)\leq 0.3$). The dotted thin circles marks the cusp region with a radius of $%
3^{\prime \prime }$ around SgrA$^{\star }$, which also contains most of the
bright emission line stars in the IRS 16 complex (Fig. 1). }
\end{figure}

\clearpage

\begin{figure}[tbp]
\plotone{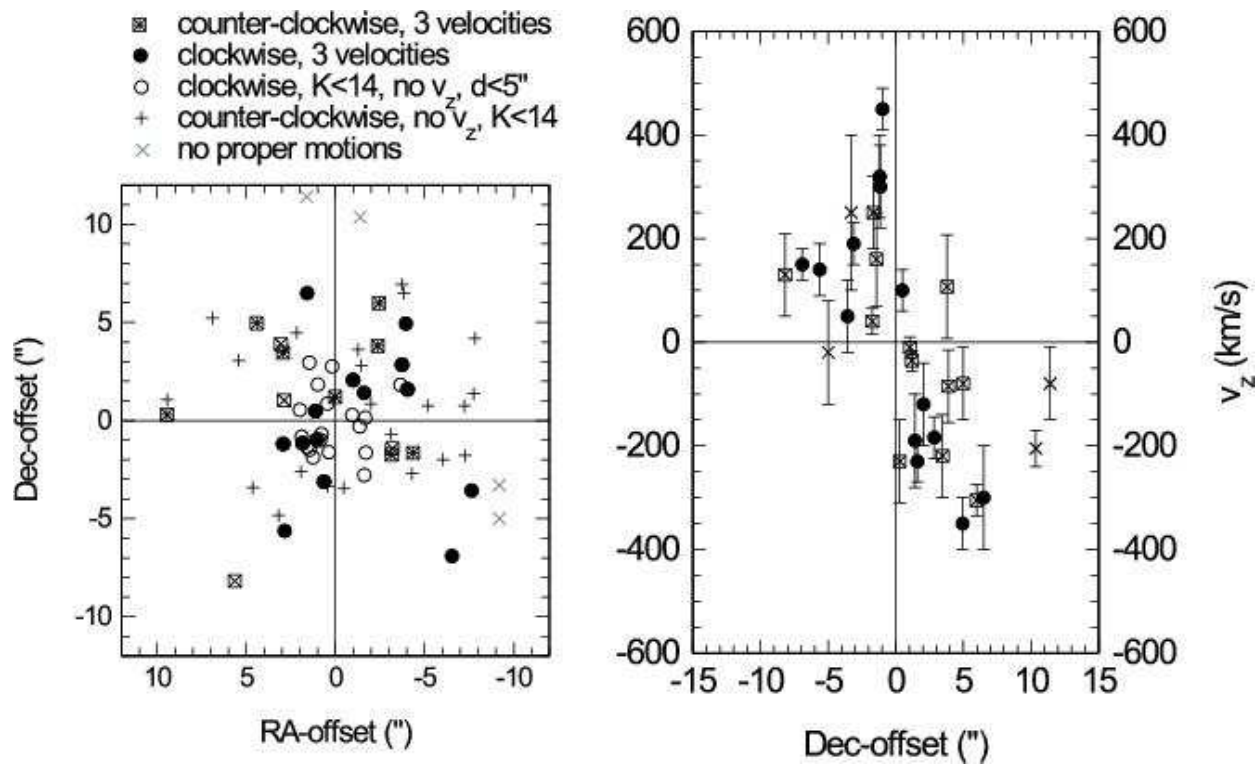}
\caption{ Left panel: Spatial distribution of early type stars. Clockwise,
spectroscopically identified stars (3 velocities) are marked by filled
circles, K$_{s}\leq14$ clockwise proper motion stars stars (2 velocities)
are marked as open circles, counter-clockwise spectroscopic stars are marked
as squares with crosses, K$_{s}\leq14$ counter-clockwise proper motion stars
are marked by plus symbols and spectroscopic stars without proper motions
are marked by crosses. Right inset: Line-of-sight velocity as a function of
declination offset from SgrA$^{\star }$, for the different types of early
type stars. Symbols are as in the left inset.}
\end{figure}

\clearpage

\begin{figure}[tbp]
\plotone{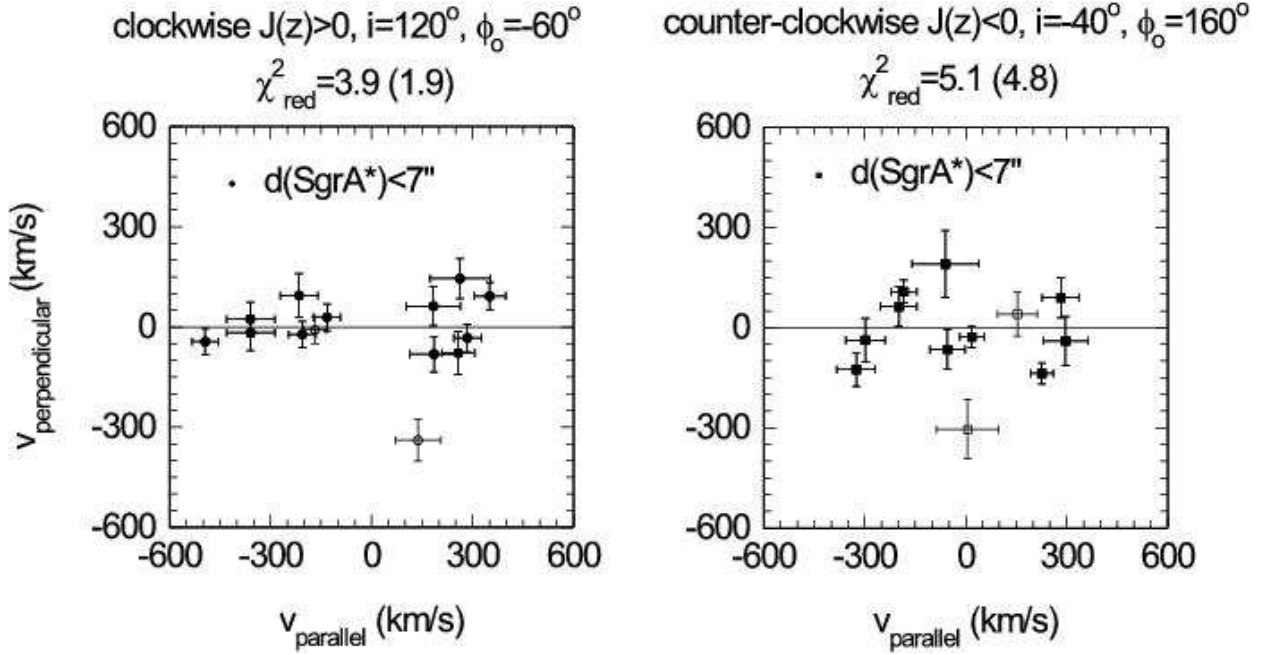}
\caption{ Velocity parallel to the best disk plane (horizontal) and
perpendicular to the disk plane (vertical) for all early type stars with
three velocities. The left inset shows stars that move clockwise, the right
inset shows stars that move counter-clockwise. The velocity data are from
Ott et al. (2003). Open symbols mark stars outside the central 7", where
proper motions are less reliable. For the clockwise stars motions are
plotted with respect to the best fitting disk plane with an inclination 120
degrees (angle relative to the plane of the sky), and with a position angle
on the sky of -60 degrees (angle on the sky, increasing east from north. For
this disk plane, the reduced chi-squared is 3.9 if all 14 stars are
included, and 1.9 if stars with p$\geq7"$ are eliminated. For the
counter-clockwise stars the best fitting plane (reduced chi-squared=5.1, and
4.8 after elimination of p$\geq7"$ stars) has an inclination of -40 degrees
and a position angle of 160 degrees. }
\end{figure}

\clearpage

\begin{figure}[tbp]
\plotone{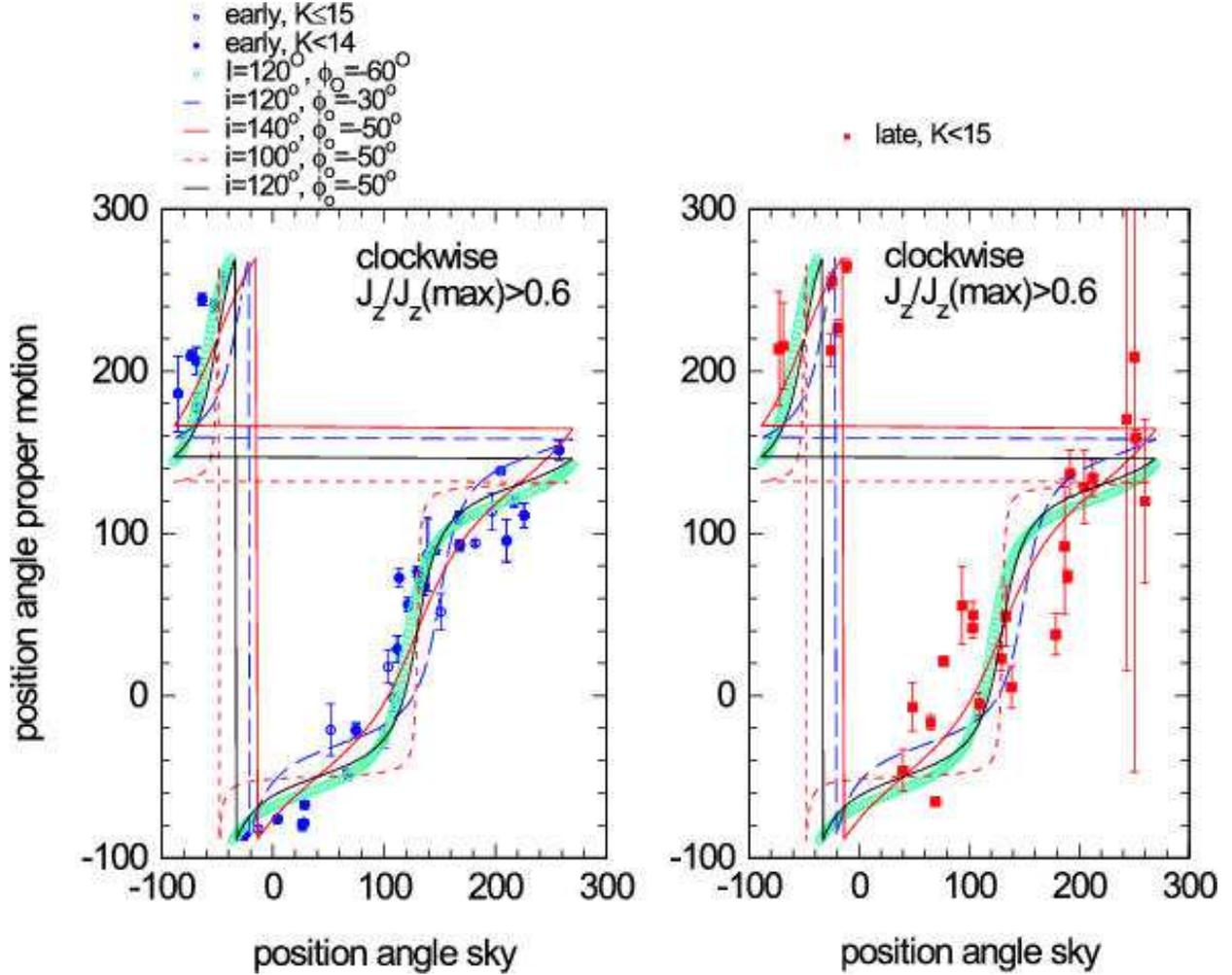}
\caption{ Left inset: Position angle on the sky (tan(x/y), horizontal) and
position angle in proper motion velocity (tan(v(x)/v(y)), vertical) for
clockwise tangential proper motion stars (filled circles K$_{s}\leq14$, open
circles K$_{s}\leq15$). The different curves (marked on the figure) denote
models of different inclination and position angle close to the best fitting
model in Figure 15. Right inset: Same but for clockwise tangential late type
stars.}
\end{figure}

\clearpage

\begin{figure}[tbp]
\plotone{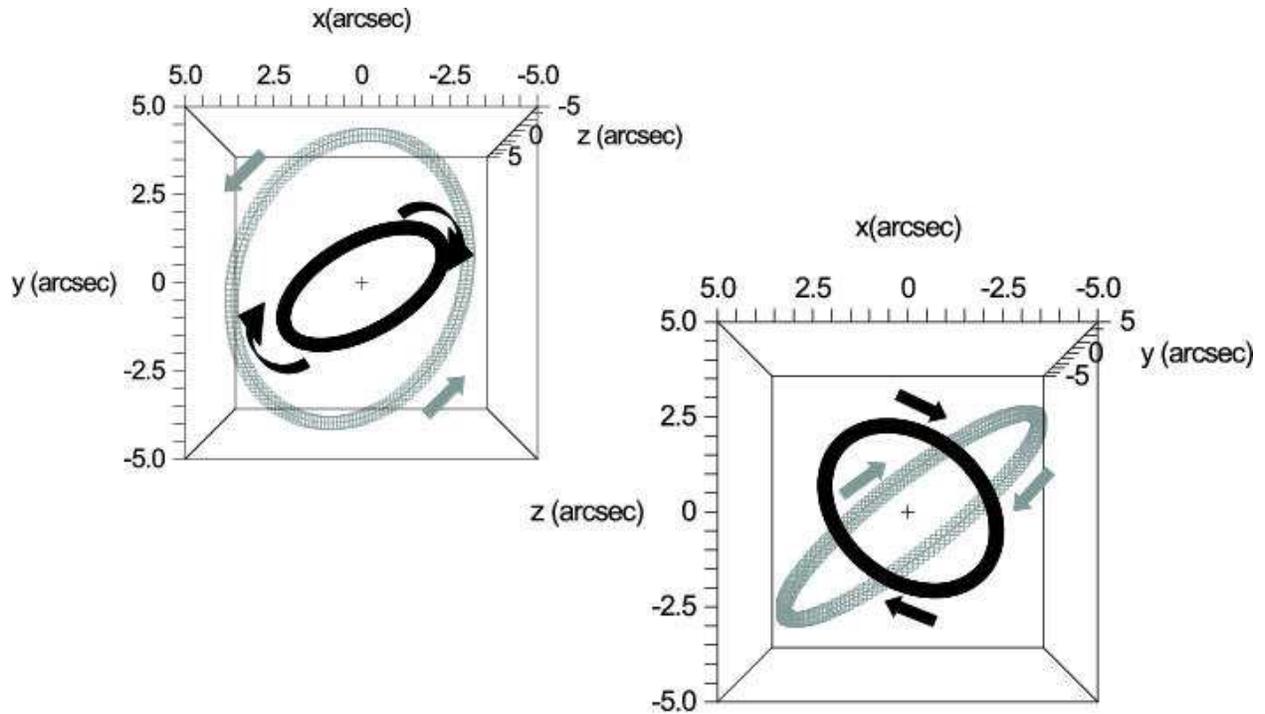}
\caption{ 3D projections of the two best fit, rotating disks and their
rotation directions. The inner thick curve denotes the characteristic radius
of the more compact, clockwise stars, while the outer, light shaded curve
 denotes the counter-clockwise stars. x and y are the coordinates on
the sky (EW, and NS) with eastward counting positive in x; z is the
line-of-sight coordinate, with the observer being located at large negative
z. In the lower right inset the top parts of both the inner and the outer
disks move toward the observer (counter to Galactic rotation).}
\end{figure}

\clearpage

\begin{figure}[tbp]
\plotone{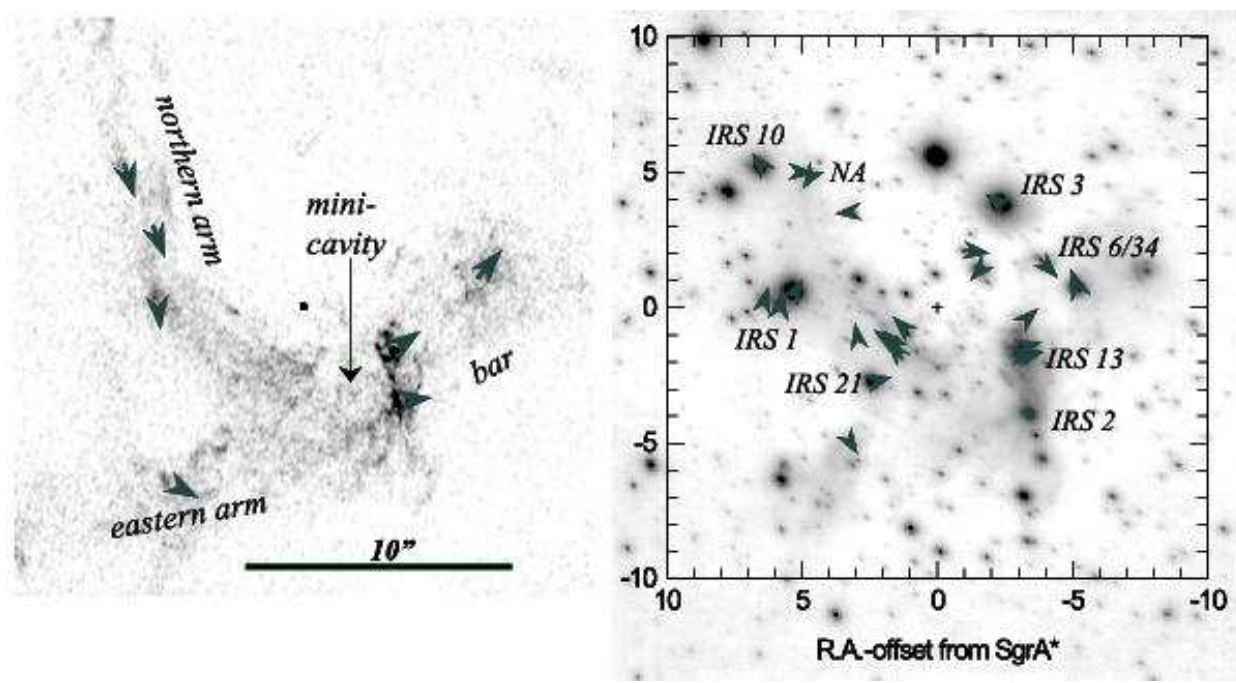}
\caption{ Comparison of proper motions of gas and stars near the gaseous
mini-spiral. Left: false color image of 1.3cm radio continuum emission from
Zhao \& Goss (1998: 0.1" resolution), with arrows denoting the proper
motions of gas components in the northern and eastern arms, and the bar plus
mini-cavity regions of the mini-spiral (Yusef-Zadeh et al. 1998). The
ionized gas in the northern arm/bar streams from the north (and behind)
anti-clockwise around SgrA$^{\star }$, and streams outward to the west and
front of the radio source. Right: NAOS/CONICA L$^{\prime }$-image
(logarithmic grey sale), with arrows denoting the proper motions of stars
with L$^{\prime }$ excess (K-L$^{\prime }$ $\geq $2). With the exception of
several dusty stars in/near IRS13, other dusty objects apparently associated
with the mini-spiral show proper motions that are distinctly different from
that of the gas. These stars likely are interacting with, and passing
through the gas and dust of the mini-spiral, but are not physically
associated with it. }
\end{figure}

\end{document}